\documentclass[fleqn,usenatbib]{mnras}
\usepackage{multirow}
\usepackage{threeparttable}

\usepackage[english]{babel}
\usepackage{newtxtext,newtxmath}
\usepackage{blindtext}
\usepackage{subcaption} 
\usepackage{graphicx} 
\usepackage{lastpage}
\usepackage[toc,page]{appendix}
\usepackage{multicol}
\usepackage[T1]{fontenc}

\DeclareRobustCommand{\VAN}[3]{#2}
\let\VANthebibliography\thebibliography
\def\thebibliography{\DeclareRobustCommand{\VAN}[3]{##3}\VANthebibliography}

\usepackage{graphicx}	
\usepackage{amsmath}	

\title[Testing the AGN Unification Paradigm]{Testing the AGN Paradigm, Part~I: a generic SED for Seyfert~1 galaxies.}

\author[Roco-Avilez et al.]{Paulina Alejandra Roco-Avilez,$^{1}$\thanks{E-mail: p.rocoavilez@ugto.mx}
Roger Coziol,$^{1}$
Juan Pablo Torres-Papaqui$^{1}$, Karla Alejandra Cutiva-Alvarez$^{1}$,
\newauthor
Mar\'{\i}a Fernanda La Rotta-Wilches$^{1}$, César David Aguirre-Gutiérrez$^{1}$ and Angélica Gabriela Sandoval-Esparza$^{1}$ 
\\
$^{1}$Departamento de Astronom\'{\i}a, Universidad de Guanajuato, Callejón de Jalisco S/N; Valenciana; C.P. 36240; Guanajuato, Gto. México.\\
}

\usepackage{float} 
\begin{document}
\label{firstpage}
\maketitle

\begin{abstract}
This article presents the first part of a study aimed at testing the unification paradigm for AGN (UPAGN) using the SED reconstruction code X-CIGALE. Our method consists in obtaining a generic SED for a large sample of Seyfert 1 (Sy1; part 1), then applying this SED to Seyfert 2 (Sy2; Part~II), expecting that the only difference will be the line-of-sight (LOS) angle, $i$, relative to the polar axis of the torus of gas and dust obscuring the broad line regions (BLRs). Our sample is composed of 3,896 Type 1, Sy1 at low redshifts, $ z<0.4$, separated into four spectral subgroups depending on the presence or absence in their spectra of narrow emission lines, Sy1N/Sy1B, and AGN wind, Sy1Bw and Sy1Nw. The generic SED produced by X-CIGALE applies to 90\% of the Sy1 in our sample. It includes a clumpy torus with an AGN engine seen face-on ($i \sim 10^\circ \pm 5^\circ$). Our analysis not only supports the existence of a torus in Sy1, in good agreement with UPAGN, but also reveals new facts about the accretion of matter and AGN wind: 1- a sudden accretion of matter from the BLR to the accretion disk triggered the wind, 2- matter from the wind replenishes the torus, consistent with a gradual formation of this structure by recurrent AGN winds, and 3- Sy1Bw and Sy1Nw eventually evolve as AGN without wind, leaving behind a torus as evidence of a higher AGN activity in their past.   
\end{abstract}

\begin{keywords}
galaxies: Seyfert -- galaxies: nuclei -- galaxies: formation -- galaxies: star formation
\end{keywords}


\section{Introduction} \label{introduction}

After the confirmation by the Event Horizon Telescope (EHT) that a super-massive black hole (SMBH) with a mass $4\times 10^6$ M$_\odot$ lays at the centre of the Milky Way \citep{2022EHTC}---a SMBH which is still actively accreting matter in a chaotic way \citep{2025Yusef-Zadeh} and was more active in the recent past \citep{2010Su,2019Heywood,2019Ponti}---the idea that the formation of any massive galaxy must necessarily pass by the formation of such object at their centre turned out to be reinforced if not vindicated. This concept is as the core of the standard model of AGN (Active Galactic Nuclei) which stipulates that all the different kinds of AGN observed, from quasar or QSO (their radio quiet counterpart, quasi stellar object) to Seyfert, LINER and radio galaxies (RGs), with various levels of emission in optical, radio and X-rays, can be explained following the same scheme, a SMBH at the centre of galaxies, accreting matter at different rates \citep{2014Heckman}. Although the link in terms of formation and evolution between the SMBHs and their galaxy hosts is still missing, as a paradigm, the standard model served as a fundamental working hypotheses that can be falsified by observations.  

As an extension of the standard model, the Unification Model of AGN (or Unification Paradigm, UPAGN) proposes that phenomenologically one can explain the two different spectral types of AGN, Type~1, with broad emission lines, and Type~2, with narrow emission lines, solely based on the orientation of the line of sight (LOS) of the observer toward the engine (the SMBH and its accretion disk): in Type~1 the engine is view face-on while in Type~2 it is viewed edge-on, such that assuming matter distributed along the LOS can absorb the emission produced by the engine, a Type~2 AGN would suffer higher extinction than a Type~1. In the case of Seyfert galaxies (Sy1 vs. Sy2), which are AGN in spiral galaxies at low redshifts \citep{1943Seyfert,1991Osterbrock}, the obscuring matter would have the form of a torus of gas and dust surrounding the engine, explaining why a Sy2 observed edge-on (LOS angle $i = 90^\circ$ relative to the polar axis of the torus) shows only narrow emission lines, while a Sy1 shows broad and narrow emission lines. Consequently, intermediate LOS angles might also be expected to explain the various intermediate Seyfert types: Sy1.5, Sy1.8 and Sy1.9 \citep{1983Osterbrock}. 

In 1983, UPAGN gained enormous support due to the important discovery made by Miller and Antonucci of broad emission lines appearing in spectropolarimetry in the prototype Sy2 galaxy NGC~1068  \citep{1983Miller}. Extending the studies to more Sy2 galaxies \citep[e.g.,][]{1992Tran}, it was then determined that the polarized broad line components were most probably produced by electrons scattered by the inner wall of the torus, assumed to be a monolithic structure where gas and dust is distributed homogeneously \citep{1992Tran}. The existence of such torus in NGC~1068 was then corroborated by \citet{1997Gallimore}, who suggested the parsec-scale structure could be an extension of the accretion disk, explaining the dependence on the LOS angle \citep[e.g.,][]{2007Tristram,2017Bisogni}. 

Revising the literature on the Unification Model in ADS,\footnote{The SAO Astrophysics Data System, which is operated by the Smithsonian Astrophysical Observatory under NASA Cooperative Agreement 80NSSC21M0056.} it is not difficult to find many observations in IR, optical and X-rays in good agreement with this paradigm \citep[see review by][]{2015Netzer}. However, as the number of observations increased, it was realized that the interpretation in terms of LOS orientation was possibly insufficient to explain the various differences in characteristics between Type~1 and Type~2 AGN. First, \citet{1994Kay} noticed that the ratio of the stellar continuum to the AGN continuum is not correlated with polarimetry in Sy2, in apparent contradiction with UPAGN. Then, \citet{2001Tran} found that many supposedly obscured Sy2s (assumed all to have large LOS angles) simply do not show broad lines in spectropolarimetry. This led him to suggest that there could be two populations of Sy2, one with hidden broad lines (HBL), the other without HBL, which he called ``pure'' Sy2 \citep{2003Tran,2011Tran}. Alternatively, some authors proposed that instead of a monolithic structure the torus could be clumpy or porous, formed by many clouds \citep{2002Nenkova,2007Tristram,2017Bisogni}. According to \citet{2011RamosAlmeida}, in Sy2 the torus is broader than in Sy1 and has more clumps with lower optical depth but higher covering factors, which implies that it is the variance in covering factor in each type that explains why some obscured Sy2s do not show HBL \citep{2015RamosAlmeida}.  

However, other authors found characteristics in Type~2 AGN that are more difficult to reconcile within UPAGN. Most specifically, it was found that the Eddington ratio, ${\rm N}_{Edd} = {\rm L}_{bol}/{\rm L}_{Edd}$, tend to be lower in Type~2 than in Type~1 AGN, suggesting that the former could have intrinsically lower accretion rates than the latter, possibly explaining the absence of broad lines in the pure Sy2 \citep{2003Nicastro,2009Panessa,2012Marinucci,2012Petrov,2015Oh,2016Pons,2020Zhao}. This has led \citet{2003Laor} to propose that because the line width in emission we observed in Type~1 AGN have a maximum, $\delta v_{max} = 25,000\ {\rm km\ s}^{-1}$, there must be a physical limit in accretion, corresponding to a luminosity ${\rm L}_{lim}$, which is required in AGN to form a BLR. Consequently, Type~2 AGN cannot form BLRs simply because ${\rm L} < {\rm L}_{lim}$. Following a similar idea, \citet{2011Trump} suggested that in Type~2 AGN the accretion process is radiative inefficient \citep[RIAF, for radiative ineficient accretion flow; also known as ADAF, for advection dominated accretion flow, see][]{2005Narayan,2014Heckman}. 

Moreover, because by definition ${\rm N}_{Edd} \propto \eta \dot{m} c^2/{\rm M}_{BH}$, i.e., the Eddington ratio depends on the ratio of the accretion rate to the mass of the BH, a difference in mass of the SMBH connected with a difference in galaxy morphology should also play a role in defining the AGN types  \citep{2006Zhang,2017Onori,2020Torres-Papaqui,2022Zhang}. Such differences were definitely observed in \citet{JP2024}, comparing the sample of 18,585 Sy2 galaxies with 4,000 Sy1 in \citet{2020Torres-Papaqui}: SMBHs in Sy2 are less massive than in Sy1, in good agreement with the lower stellar masses of their host galaxies, and tend to have later-type morphologies and higher SFRs than the Sy1 host galaxies. However, despite their lower BH masses, \citet{2020Torres-Papaqui} confirmed that ${\rm N}_{Edd}$ is lower on average by 0.8 dex in Sy2 than in Sy1, which, coupled to an average lower AGN luminosity, imply they are definitely accreting matter at lower rates. 

Howbeit, finding such physical differences between Sy1 and Sy2 is possibly not enough to falsify UPAGN, since these differences might either be consistent with differences in the tori structures or related to some kind of evolution between the two types. More specifically, a difference of morphology and SFR could engender differences in torus covering factor and/or optical depth  \citep{2016Mateos, 2017Audibert}, or considering the small sizes of the tori (a few parsecs), there could also exist a connection between these structures and outflows \citep[OFs or AGN winds; e.g.,][]{2006Elitzur, 2012Wada}. Moreover, an OF structure instead of a clumpy torus would possibly be more adequate to explain observations of polar emission in AGN \citep{2014Schartmann,2016Honig,2016Wada,2017Stalevski,2018Leftley,2020Williamson}. Finding any physical connection between OFs and tori would have a major impact on our understanding of AGN winds and their roles in the formation/evolution of galaxies. More concretely, if the accretion of matter at low luminosity switches from radiative with OF to radio Jet without OF, as the ADAF hypothesis suggests, then BLRs could be expected to eventually disappear as the AGN luminosity goes down \citep{2006ElitzurShlosman}, leading to a possible evolution from unobscured Type~1 to obscured Type~2, and maybe thereafter to LINER-like or normal-like spiral galaxies with starving SMBHs at their centre \citep{2018DiPompeo,2022Ricci}. 

In our previous study of a large sample of AGN with different spectral classifications \citep{JP2024}, we did report a decrease in the frequency detection of OF in low-luminosity AGN (RG-AGN/LINER/Sy2) compared to high-luminosity AGN (Sy1), which apparently could be consistent with different accretion characteristics (ADAF in the first group and radiative in the second group). However, comparing Sy1 with Sy2 after correcting for the higher stellar contribution in the continuum of the latter, we did not discern a difference in AGN power law, suggesting that the accretion process in these two types are similar, more precisely, both are in a radiative mode, just accreting at different rates due to a difference in BH mass. Consistent with this interpretation, we also found no difference in AGN wind intensity between the two Seyfert types, except, in general (not only in the Seyfert galaxies), in any AGN detected in radio, where OFs are systematically more intense, suggesting that both radiative and jet mode in these galaxies could be simultaneously active. In conclusion, the only differences we observed between Seyfert galaxies with and without OF are differences in morphology, accretion rate (or AGN luminosity) and SFR: Seyfert with OF are in later-type galaxies, with higher accretion rates and star formation rate (SFR). Since no evolution from Sy1 to Sy2 is possible, because of their differences in morphology and BH mass, this must imply that these two types of AGN must have followed different evolutionary paths related to different formation processes, where an OF characteristically appear at an earliest phase of their evolution, when the galaxies are richer in gas.

On the other hand, our two previous studies of OF in AGN revealed almost nothing about the torus and consequently cannot allow us to establish whether the differences we observed are consistent with UPAGN. To explore this question further, we decided to apply the spectral energy distribution (SED) reconstruction code X-CIGALE \citep{boquien19,yang22} on our large samples of Sy1 and Sy2 to test if the conditions of UPAGN are verified. In this first paper we concentrate on the Sy1 testing the following two hypotheses: 1- the SED is generic, applying to all the galaxies in our sample, and 2- the LOS angle relative to the polar axis of the torus is small, consistent with an unobstructed view of the engine. 

\section{Sample and Data} \label{sample and data}

The sample of 3,896 Sy1 we are using in this study comes from the sample of Sloan digital sky survey (SDSS) galaxies with redshift $z<0.4$ defined by \citet[][]{2020Torres-Papaqui} for their study of OF in AGN. The sample is divided into four spectral subgroups: Sy1N, where both narrow and broad line components of H$_{\beta}$ appear in the spectra, Sy1B, where only a broad component appear, and Sy1Nw/Sy1Bw, where OFs are detected in the narrow line [OIII]$\lambda5007$. In Table~\ref{tab:Sy1}, the number of galaxies in each of the four spectral subgroups is shown. Comparing the values in the first column (TP20), one can see that the number of galaxies decreases along the sequence Sy1N$\rightarrow$Sy1Nw$\rightarrow$Sy1B$\rightarrow$Sy1Bw. Therefore, two components in emission is the norm in Sy1, and in both spectral subtypes AGN winds are equally frequent---OFs being very common in Sy1 \citep{2020Torres-Papaqui}.

\begin{table}
\caption{Number of galaxies in each spectral subgroup in the two samples, SW and SW2M; the percentages are relative to the total in each column, while in the last line, the percentages are relative to the total in column TP20 from data in \citep{2020Torres-Papaqui}.}
\label{tab:Sy1}
\begin{center}
\begin{tabular}{| c | c | c | c | }
\hline
Samples & TP20 & SW & SW2M  \\ 
\hline
Sy1N & 1534(40\%) & 1126(37\%) & 833(37\%)  \\ 
Sy1Nw & 974(25\%) & 780(26\%) & 551(25\%) \\ 
Sy1B & 905(23\%)  & 701(23\%) & 503(22\%) \\ 
Sy1Bw & 483(12\%) & 431(14\%) & 350(16\%) \\ 
\hline
Total & 3896 & 3038(78\%) & 2237(57\%) \\ 
\hline
\end{tabular}
\end{center}
\end{table}

To construct the SED of these AGN using X-CIGALE, what is needed is photometric data of high quality covering the largest range of wavelengths possible. For our analysis, all the data were obtained by cross-correlating the list of Sy1 in \citet{2020Torres-Papaqui} with entries in different catalogues available in VIZIER\footnote{at https://vizier.cds.unistra.fr/viz-bin/VizieR} using the option ``MATCHES'' in TOPCAT.\footnote{TOPCAT is an interactive graphical viewer and editor for tabular data capable of handling large and sparse datasets with correlation functionality; 	https://www.star.bris.ac.uk/~mbt/topcat/} From SDSS, we retrieved photometric data in the optical for all 3,896 galaxies at five wavelengths, 354 nm (u), 476 nm (g), 628 nm (r), 769 nm (i) and 925 nm (z).

From WISE, we retrieved another four photometric data in MIR/FIR for 3,038 galaxies, centred at 3.4 µm (W1), 4.6 µm (W2), 12 µm (W3) and 22 µm (W4). In the WISE catalogue the quality of each flux is indicated by a flag, \textit{qph}, taking four values, A, B, C, and U: flag~A corresponds to a signal to noise ratio S/N $<10$, flag~B to the range $3 \leq$ S/N $\leq 10$, flag~C to $2 \leq$ S/N $\leq 3$, while flag~U indicates S/N\ $ <2$. Without applying any pre-selection criterion in photometry, 98\% of the galaxies in our sample in the bands W1 and W2 have flag~A while the rest have flag~B. For the bands W3 and W2, the flags are either B (in majority) or A, with only a few C (no more than 18\%). 

Finally, for 2MASS we could find three photometric data in NIR for 2,237 galaxies, with wavelengths centred at 1.235 µm (J), 1.662 µm (H) and 2.159 µm (K$_{s}$). Since adding new wavelength bands reduces the number of objects (from SDSS to WISE to 2MASS), we decided to study the Sy1 by forming two samples: one containing only SDSS-WISE data (SW) the second with SDSS-WISE-2MASS data (SW2M). This was done with the idea of determining how adding data in different bands affects the SED obtained by X-CIGALE. 

In Table~\ref{tab:Sy1}, we can verify that after cross-correlations the number of SDSS galaxies decreases to 78\% in the SW and 57\% in the SW2M, while the relative frequencies of each spectral subgroup decrease the same way as in the total sample (TP20; comparing the percentages in each column).

\subsection{Transformation of magnitudes into fluxes} \label{transformation of magnitudes into fluxes}

The next step in our preparation of the data consists in transforming the magnitudes in fluxes. For SDSS, we used Equation~\ref{SDDS-flujo}, which is the hyperbolic magnitude defined by \citet[][]{1999Lupton}, where $f_{0}$ is the flux of an object with conventional magnitude zero, and $b$ is a softening parameter typical of 1~$\sigma$ noise level for the seeing in an aperture with PSF of 1~arcsec. The values of $b$ and $f_{0}$ in each band of SDSS filters can be found in one of the web pages of \citet{SDSSwebsite}.

\begin{equation} \label{SDDS-flujo}
f (Jy) = 2 b f_{0} \sinh\left(\frac{m \ln(10)}{-2.5}-\ln({b})\right)
\end{equation}

The uncertainty on the fluxes are calculated by applying Equation~\ref{SDDS-incertidumbreflujo}, where \textit{dm} is the uncertainty on the magnitude. 
\begin{equation} \label{SDDS-incertidumbreflujo}
df (Jy) = \left|\frac{(f dm/[-2.5/\ln(10)])}{\tanh(\{m/[-2.5/\ln(10)]\}-\ln(b))}\right|
\end{equation}

To transform the magnitudes in the WISE catalogs, we used Equation~\ref{WISE-flujo}, where $F_{\nu_0}^{*}$ is the zero magnitude flux density derived for sources with a power-law spectra, $F_{\nu}$ $\propto$ $\nu^{-\alpha}$, and $f_{c}$ is a color correction. 
\begin{equation} \label{WISE-flujo}
f (Jy) = (F_{\nu_0}^{*}/f_{c}) \times 10^{(-m/2.5)} 
\end{equation}
In \citet{flux-WISE} one can find the possible values for different exponent $\alpha$ in Table~1 and typical colors they produce in Table~2, after applying the corresponding color correction. After comparing the colors of the Sy1 in our SW sample with those in Table~2, we opted to use the power law $F_{\nu}$ $\propto$ $\nu^{-1}$, where $f_{c}$ is equal to 0.9921, 0.9943, 0.9373 and 0.9926, for the band W1, W2, W3, and W4, respectively.

Equation~\ref{2MASS-flujo} was used for 2MASS, where $ F_{\nu_0}$ is the zero magnitude in each band as listed in Table~2 of \citet{flux-2MASS}.
\begin{equation} \label{2MASS-flujo}
f (Jy) = F_{\nu_0}\times 10^{(-m/2.5)} 
\end{equation}

The uncertainties for the WISE and 2MASS fluxes were estimated by determining the differences between the fluxes calculated and the fluxes incremented $m\pm dm$, where $dm$ correspond to the uncertainties in magnitude for each measured as reported in the catalogues of each survey. 

\subsection{Peculiar distribution in redshift} \label{Peculiar distribution in redshift}

\begin{figure*}
\begin{multicols}{3}
    \includegraphics[width=\linewidth]{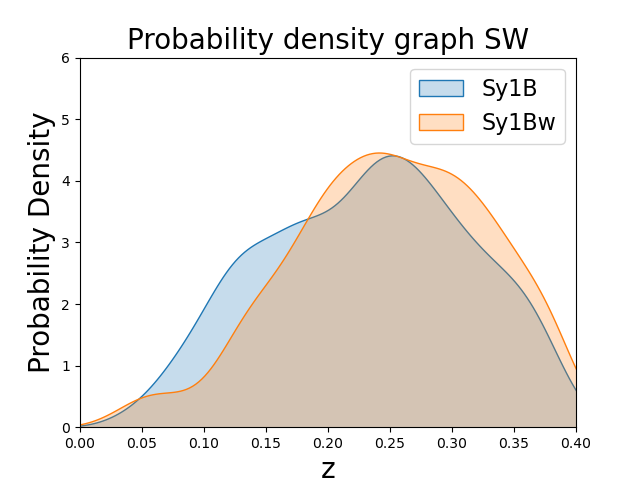}\par 
    \includegraphics[width=\linewidth]{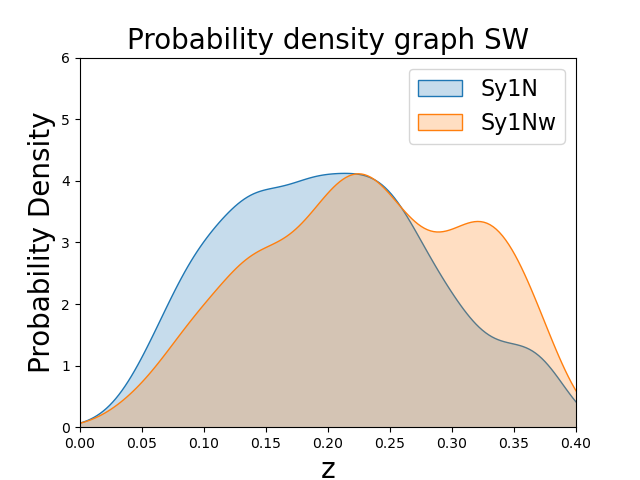}\par
    \includegraphics[width=\linewidth]{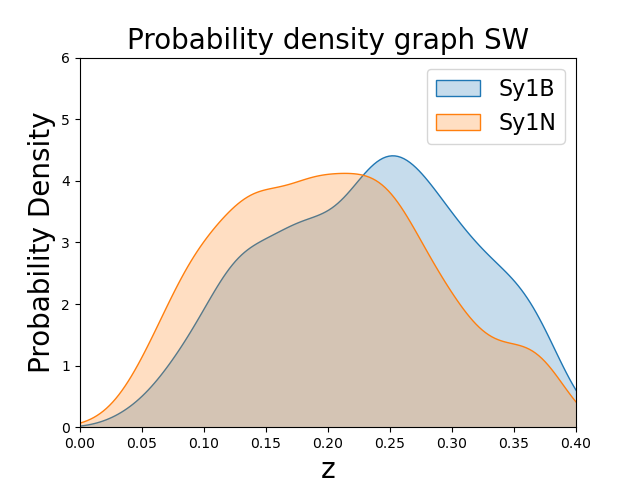}\par
    \end{multicols}
\begin{multicols}{3}
    \includegraphics[width=\linewidth]{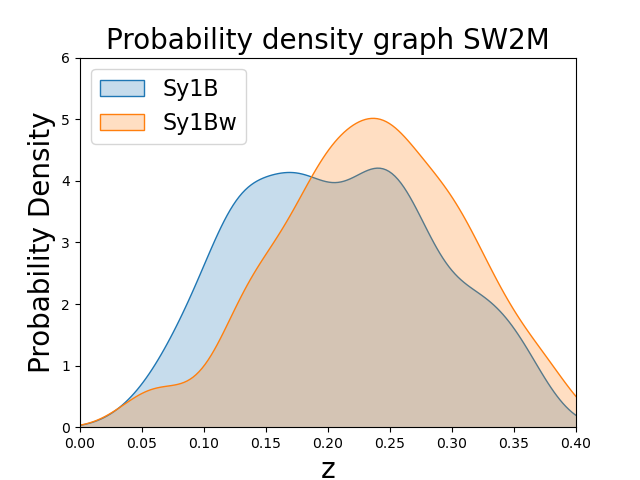}\par
    \includegraphics[width=\linewidth]{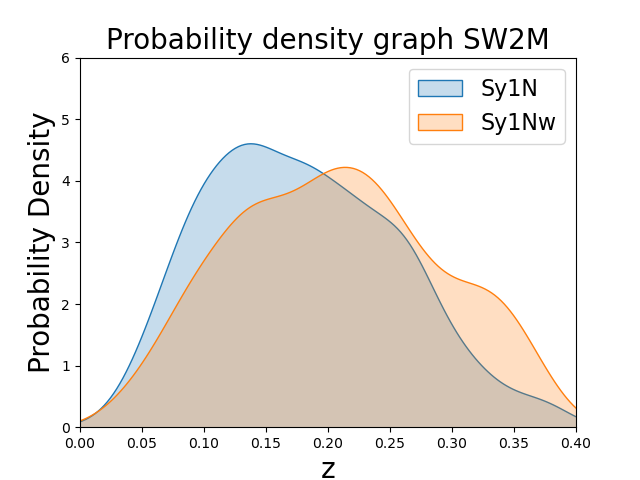}\par
    \includegraphics[width=\linewidth]{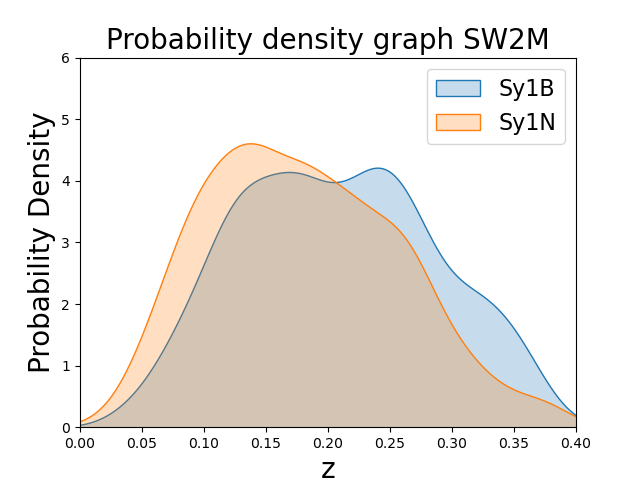}\par
\end{multicols}
\caption{Density distributions in $z$ in the four spectral subgroups in the two samples SW and SW2M. }
\label{fig:z-Sy1}
\end{figure*}

In Figure~\ref{fig:z-Sy1}, we present in the two samples of Sy1, SW and SW2M, the density plots for the distributions in redshift ($z$) in each of the spectral subgroup. In each sample, Sy1B tend to be found at higher $z$ than Sy1N and Sy1 with AGN wind tend to be located at higher $z$ than their counterparts without wind \citep{2020Torres-Papaqui,JP2024}. In Table~\ref{tab:resultado-Dunn-z} we show the results of applying a Kruskal–Wallis, nonparametric, ANOVA test, with Dunn’s multiple comparisons tests, to compare the data. A detailed description of these tests can be found in page 4 and 5 of \citet{2020Torres-Papaqui}, together the description of the different levels of significance in their Table 1. These tests confirm at the highest level of significance the differences observed in redshift (except between the Sy1B and Sy1Nw).

\begin{table}
\centering
\begin{threeparttable}
\caption{Summaries of Dunn’s Multiple comparisons test for the density distribution in $z$. }
\label{tab:resultado-Dunn-z}
\begin{tabular}{|l|c|c|c|}
\hline
Pairs compared & SW & SW2M & higher $z$\\
\hline
Sy1B-Sy1N   & **** & **** & Sy1B \\ 
Sy1B-Sy1Bw  & ***  & **** & Sy1Bw \\ 
Sy1N-Sy1Nw  & **** & **** & Sy1Nw \\
Sy1Bw-Sy1Nw & **** & **** & Sy1Bw \\ 
Sy1Bw-Sy1N  & **** & **** & Sy1Bw\\ 
Sy1B-Sy1Nw  & $ns$ & $ns$ & $ns$ \\ 
\hline
\end{tabular}
\begin{tablenotes}
      \item Notes: Tests performed on spectral subgroups with last column identifying which subgroup of the pair is at higher $z$; levels of significane, ($****$ , $p < 0.0001$), ($***$, $p < 0.001$), ($**$, $p < 0.01$), ($*$, $p < 0.05$), ($ns$, $p \ge 0.05$), which is non significative.
    \end{tablenotes}
\end{threeparttable}
\end{table}

Since it is slightly easier to detect broad emission lines at higher redshifts, one might suggest that the higher number of Sy1B at high $z$ in our samples is due to an observational bias. However, this would not explain the apparent deficit of Sy1B at low $z$. Consequently, the difference in redshift is most probably genuine, although it is not clear what this could mean physically or how this could be interpreted within UPAGN. On the other hand, no observational bias is expected to affect the detection of OF, since the separation in spectral subgroup is based on H$\beta$ lines, while the detection of OF is based on the detection of a blue shift component of the narrow line [OIII]$\lambda 5007$, two different, unreated . 

Beside, there are other physical differences between the Sy1 in the different spectral subgroups that could explain what we observe. For example, a difference in morphology of the host galaxies: since Sy1B/Sy1Bw tend to be in earlier type spirals than Sy1N/Sy1Nw \citep[][]{2020Torres-Papaqui}, a more massive and luminous bulge might make them easier to detect them at high $z$. 
On the other hand, due to the downsizing phenomenon (massive galaxies forming before less massive galaxies), Sy1B having more massive BHs than Sy1N, early-type Sy1B would naturally be expected to be at higher $z$ than late-type Sy1N. In \citet{JP2024} it was concluded that the differences in redshifts between RG/Sy1 (high $z$) and LINER/Sy2 (low $z$), are genuine, and can be explained by alluding to different states of formation/evolution of the galaxy hosts and their SMBHs. If this is the case, differences in the tori of Sy1B and Sy1N might also be expected, which would allow, possibly, to better understand the phenomenological origin of UPAGN.   

\section{Method of analysis} \label{method of analysis}

According to UPAGN, the only difference expected between Sy1 and Sy2 is the LOS angle, $i$, relative to the tori of gas and dust surrounding the engine. In Sy1, the engine is expected to be directly visible, which implies that $i$ must be small, close to $0^\circ$, relative to the polar axis of the torus. As a corollary for the Sy1, the engine is also expected to be the same in the spectral subgroups, and, therefore, to all have the same SED. Consequently, our method to test UPAGN in Sy1 using X-CIGALE \citep{boquien19, yang22} is relatively straightforward \citep{2009Gaskell}: it consists in 1- testing whether the SED is generic, that is, one model applies to all the Sy1, and 2- verifying that the LOS angle $i$ is relatively small, close to $0^\circ$. 

\begin{table*}
\centering
\caption{Summary of inputs used in \textsl{X-CIGALE} for ultimate run of Sy1 generic SED model with polar emission.}
\begin{tabular}{||c c c||} 
 \hline
Parameters & Values & Descriptions  \\ 
 \hline\hline
 \multicolumn{3}{||c||}{\textbf{sfhdelayed}} \\ 
 $\tau_{main} $ & 7000 &  e-folding time in Myr. \\
 $age_{main}$ & 7000 & Age of the oldest stars (Myr). \\ \hline
 \multicolumn{3}{||c||}{\textbf{SSP:bc03}} \\ 
 $Z$  & 0.02 &  Metallicity. \\
Separation Age  & 10 &  Age difference between the youngest \\ && and the oldest stellar populations (Myr). \\ \hline
 \multicolumn{3}{||c||}{\textbf{Nebular emission.}} \\ 
 $logU$ & -2.5 & ionization parameter. \\
 $z_{gas}$ & 0.02, 0.03 & Metallicity of the gas.\\
 $n_{e}$ & 1000 & Electron density. \\
 $lines_{width}$ &  3000 & Line width (km/s). \\ \hline
 \multicolumn{3}{||c||}{\textbf{Dustatt\_modified\_CF00}} \\ 
 $Av_{ISM}$& 1.0 & Attenuation of the V  band in the interstellar medium. \\
 $slope_{ISM}$ & -0.7 & Slope of the power law of attenuation in the ISM. \\ \hline
 \multicolumn{3}{||c||}{\textbf{SKIRTOR}} \\
 $pl$ & 1.0 & Power law exponent that establishes \\ && the radial gradient of the dust density. \\
$R_{max}/R_{min}$ & 30 & Relationship between the maximum \\ && and minimum radii of the dust torus. \\
\textit{oa}&  30, 40, 50, 60, 70, 80 & Opening angle. \\
\textit{i} & 0, 10, 20  & Viewing angle. \\
$f_{AGN}$ & 0.1, 0.2, 0.3, 0.4, 0.5, 0.6, 0.7, 0.8, 0.9 & Fraction of AGN contribution to the IR luminosity.\\ \hline
\multicolumn{3}{||c||}{\textbf{Polar emission}} \\
 E(B-V) & 0.1, 0.3, 0.5 & Extinction in the polar direction in magnitudes.\\
 temperature & 100, 200 & Polar dust temperature in K.\\ \hline
\end{tabular}
\label{tab:modelo_cigale}
\end{table*}

In X-CIGALE, the SED is constructed by determining the physical parameters defining the AGN and its torus, emitting in UV/Opt and MIR, and of the stellar populations which contribute mostly in Opt/NIR, with possible contributions in UV and FIR depending upon the level of star formation (characterized by the star formation rate, SFR), with reemission by dust in MIR/FIR and dust extinction in UV/Opt. The different modules we used in X-CIGALE to construct a SED are given in Table~\ref{tab:modelo_cigale}, together with the final values that were adopted to be tested for the generic SED. 

The first module characterizes the star formation history (SFH). Because Seyfert galaxies are spirals, we first started with the exponential function, \textit{sfh2exp}, varying the e-folding time, $\tau_{main}$, from short, burst-like episodes, favouring a rapid formation of the bulge, to long lasting episodes, consistent with constant star formation (SF) in the disks. However, since Sy1 galaxies tend to be in early-type spirals, we also tried a SFH described by the \textit{sfhdelayed} function with various  $\tau_{main}$ (short and long). 

To test the different solutions, we used two criteria, the $\chi^{2}$ of the best fits, which should be as close to one as possible, and the Bayesian Information Criterion (BIC), BIC$ = \chi^{2}+k \ln(n)$, which also take into account the number of free parameters in the modules that can be fitted, $k$, and the sum of the photometric bands in the samples, $n$; where $n=9$ in SW and $n=12$ in SW2M. Comparing different fits, the model with smallest BIC is accepted as the best one. 

Applying these two criteria, the best model for the generic SED turned out to be described by the \textit{sfhdelayed} function with a relatively long e-folding time, $\tau_{main}= 7,000$ Myr, and relatively high (starburst-like) SFR. This solution seems to be in good agreement with a scenario of early-type spiral galaxies (with massive bulges) that are still actively forming stars in their disks; incidentally, the SFRs produced by \textit{sfhdelayed} were found to be slightly lower than when \textit{sfh2exp} was used. The age of the oldest stellar populations, varying from young (1,000 Myr) to old (10,000 Myr), was found to converge to 7,000 Myr in the generic SED, consistent with an epoch of formation at ~$z\ge 1$. 
\begin{figure}
	\includegraphics[width=0.8\columnwidth]{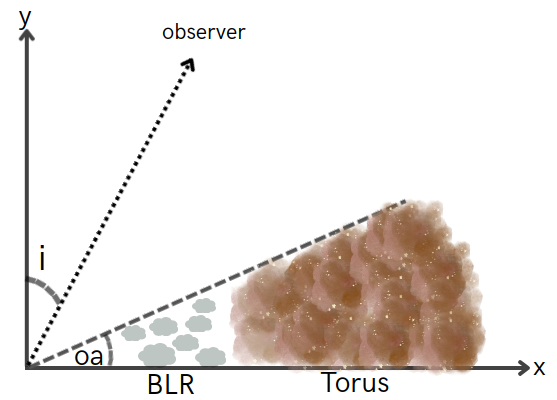}
    \caption{Illustration of the angles as defined in X-CIGALE: $i =$ LOS angle (when $i = 0$, torus viewed face-on) and $oa$ = half-opening angle. In \citet{2012Stalevski}, $i = \theta$, $oa = \Theta/2$. Torus and BLR not to scale.}
    \label{fig:model-oa}
\end{figure}
For the stellar populations, the Simple Stellar Population (SSP) model of \citet[][bc03]{bc03} was employed, adopting a solar metallicity and a small difference in age between the old and young stellar populations, consistent with continuous SF. For the stellar attenuation (extinction) by dust, the module dustatt\_modified\_CF00 which is based on the study of \citet{charlot} was utilized. The nebular emission module was also activated, but since the version of X-CIGALE we used did not include AGN emission, we fixed the ionization factor and electron density to typical values in star forming regions \citep{2006Osterbrock}. However, we did put the maximum FWHM available, $3,000\ {\rm km s}^{-1}$, which is more typical of AGN than of SF regions. 

Note that we did not experiment exhaustively with the modules of stellar population, dust extinction and nebular emission, because our main goal, testing UPAGN, suggested we concentrate on the parameters that could have a critical impact, which are the AGN contribution to the SED and its relation with the torus. This was done using the module SKIRTOR, which is a model of a clumpy torus including the AGN contribution \citep{SKIRTOR}.The geometry of the torus is shown in Figure~\ref{fig:model-oa}. After fixing the values of the gradient in dust density and the ratio of maximum and minimum radii of the torus according to \citet{fritz}, we first ran X-CIGALE using small values of $i$ leaving $oa$ and the fraction of AGN, $f_{AGN}$, to vary over the largest range possible. The best fit SEDs rapidly converged to small angle values in the majority of the Sy1 galaxies in the two samples. To double check this result, we also tried different runs with only larger $i$. In SKIRTOR, $i$ can vary from $0^\circ$ (face-on for no HBL) to $90^\circ$ edge-on (for HBL), however, we could verify using the $\chi^{2}$ and BIC that the best fits for the largest number of galaxies in our samples were always favouring low $i$. As a final test, we reduced the range of values of $i$ as shown in Table~\ref{tab:modelo_cigale}, and ran X-CIGALE a last time including polar emission. 

\subsection{Applying X-CIGALE in detail: testing the Sy1 generic SED}

To determine the values of the SFH and SKIRTOR modules for the generic SED, we first separated the four spectral subgroups of Sy1 in height redshift bins, then, selected in each bin one galaxy with the highest quality fluxes in WISE as representative of the whole bin. The number of galaxies in each bin, in both samples, are reported in Table~\ref{tab:BIN_SW}. Adopting the Sy1B spectral subgroup as prototype for the generic SED, we ran X-CIGALE on each representative galaxies in the bins. We started with the SFH \textit{sfh2exp} with small e-folding time, $\tau_{main}$, using limited ranges of \textit{i}, but leaving the other parameters, $oa$ and $f_{AGN}$, to vary over the largest range of values possible, and each time testing, comparing the best fits.

\begin{table}
\begin{center}
\caption{Redshift intervals for analysis with X-CIGALE, counting galaxies in each bin in the SW and SW2M samples.}
\label{tab:BIN_SW}
\centering
\begin{tabular}{| c | c | c | c | c |}
\hline
 & \multicolumn{4}{c}{\# of galaxies in SW/SW2M}\\
 \cline{2-5}
$\Delta z$ & Sy1B & Sy1Bw & Sy1N & Sy1Nw \\ 
\hline
$[0 - 0.05)$ & 6/6  & 4/3 & 18/17 & 10/10 \\ 
$[0.05 - 0.10)$ & 35/35 & 13/13 & 124/120  & 55/54 \\ 
$[0.10 - 0.15)$ & 101/98 & 34/34 & 211/197 & 106/99 \\ 
$[0.15 - 0.20)$ &112/102  & 67/64 & 223/179 & 124/105 \\ 
$[0.20 - 0.25)$ & 146/107 & 90/83 & 233/151 & 167/118 \\ 
$[0.25 - 0.30)$ & 143/80 & 103/87 & 168/112 & 113/73\\ 
$[0.30 - 0.35)$ & 99/56 & 75/46 & 86/39 &  144/68 \\ 
$[0.35 - 0.40)$ & 59/19 & 45/20 & 63/18 & 61/24  \\ 
\hline
\end{tabular}
\end{center}
\end{table}

X-CIGALE provides two output values: the first is identified as "best", which corresponds to the model with the best fit, i.e., the one with the lowest $\chi^{2}$ value, and the second is "bayes," which takes into account the weights of all models. In addition, the program generates a plot of the best fits with residual, allowing for a direct examination of the resulting SEDs. Obtaining high values of $\chi^{2}$ and BIC were interpreted as indicating we need to adjust the module values or change the ranges of these values, determining by eye examination of the best fit SEDs and residual which parameters in each module needed to be modified. This is how we determined that the \textit{sfhdelayed} function with long $\tau_{main}$ yields better results than the \textit{sfh2exp} (with any $\tau_{main}$). We also determined that the best value for the LOS angle was close to $i = 10^\circ$. Constant comparison of the best fit SEDs revealed that $f_{AGN}$ increases using \textit{sfhdelayed} as compared to \textit{sfh2exp}, while SFR simultaneously decreases.  

\begin{figure}
\includegraphics[width=\columnwidth]{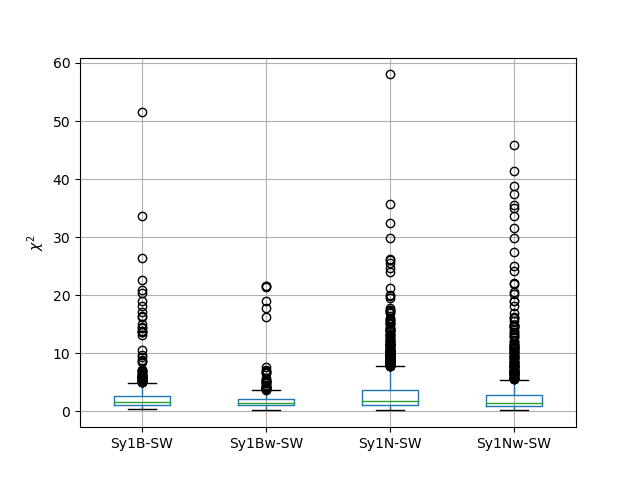}
\includegraphics[width=\columnwidth]{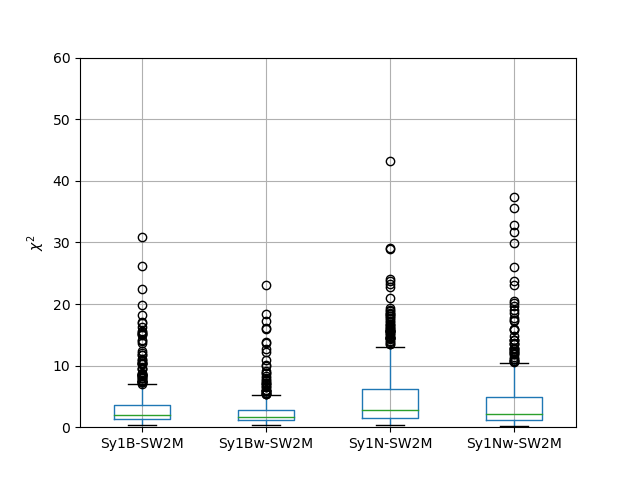}
\caption{Box-whisker plots of $\chi^2$ applying the generic SED for Sy1 on galaxies in the samples SW (upper panel) and SW2M (lower panel). The sizes of the boxes are determined by the quartiles, Q1, Q2 and Q3, respectively, $q_n(0.25)$, $q_n(0.50)$ and  $q_n(0.75)$, the whiskers being at $\pm 1.5\times IRQ$, where $IRQ = Q3-Q1$, the inter quartile range. All data with values lower/higher that the whiskers are considered outliers.}
\label{fig:Sy1-chi}
\end{figure}

Finally, we tried a solution with polar dust, which in X-CIGALE \citep{yang22} is modelize by adding dust extinction within the aperture of the torus (with angle $1-oa$) under the form of a color excess E(B-V) and dust temperature. The final values tested are those indicated in Table~\ref{tab:modelo_cigale}. Although the differences in $\chi^{2}$ and BIC were marginals, after comparing visually the SEDs with and without polar dust it was determined that polar emission on average improves the fits in the IR range (w4 of WISE). This is why we decided to include this component in our generic SED for Sy1. However, we do not claim this result is a definitive support for polar dust in Sy1 galaxies. 

\begin{table}
\centering
\caption{Comparing the medians $\chi^{2}$ of the best fits in the SW and SW2M samples.}
\label{tab:chiSy1}
\begin{minipage}{.5\linewidth}
\centering 
\begin{tabular}{|c|c|c|c|c|}
\hline
Subgroups &  (SW) & (SW2M)  \\
& $\chi^{2}$&$\chi^{2}$\\
\hline
Sy1B  & 1.67 & 2.01 \\ 
Sy1Bw & 1.50 & 1.71 \\ 
Sy1N  & 1.74 & 2.86 \\ 
Sy1Nw & 1.45 & 2.09 \\ 
\hline
\end{tabular}
\end{minipage} 
\end{table}

\begin{figure}
\includegraphics[width=\columnwidth]{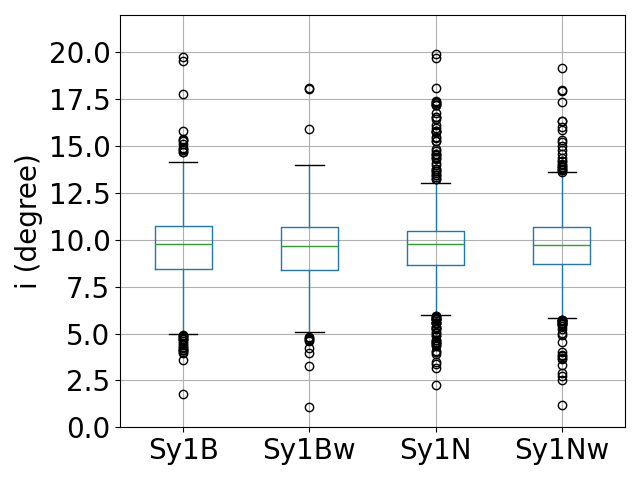}
\caption{Box-whisker plots for \textit{i} as produced by the Sy1 generic SED on galaxies in the sample SW.}
\label{Sy1i_BxPl}
\end{figure}

The same procedure as described above for the Sy1B was applied to the Sy1N and the Sy1 with AGN wind, concluding that the parameters in Table~\ref{tab:modelo_cigale} can be applied to all these galaxies. As a final test, we ran X-CIGALE on the ensemble of galaxies in the sample SW. In Figure~\ref{fig:Sy1-chi} we present the box plots of the $\chi^{2}$ for the best SED fits, including outliers. In the upper panel for SW, we count 52 out of 701 (7.2\%) outliers in the Sy1B and 121 out of 1126 (10.7\%) in the Sy1N. Although there is a slight increase in the proportion of outliers in Sy1N, this difference seems insignificant given that the number of galaxies in this subgroup is larger. For the Sy1 with AGN winds, the numbers of outliers are very similar: 34 out of 431 (7.8\%) in the Sy1Bw and 109 out of 780 (13.9\%) in the Sy1Nw. Eliminating the outliers, the medians for the $\chi^{2}$ for the whole sample of Sy1 as reported in Table~\ref{tab:chiSy1} show values well below two. 

In Figure~\ref{Sy1i_BxPl}, we also present the box-whisker plots for the \textit{i} Bayes values. From the medians in Table~\ref{tab:SFR-fagn-i-oa}, we conclude that \textit{i} stays close to $9.7^\circ/9.8^\circ$ in the four spectral subgroups. Applying a Kruskal-Wallis test yields a p-value equal to 0.95, implying that there is no significant difference in \textit{i} between the Sy1 subgroups. In other words, in all these Sy1 galaxies the torus is seen face-on, leaving an unobstructed view of the engine. In general, therefore, we conclude that the solution we have determined using X-CIGALE can be accepted as a generic SED consistent with UPAGN for the whole ensemble of Sy1 with and without AGN wind. 

\begin{figure*}
\includegraphics[width=0.8\columnwidth]{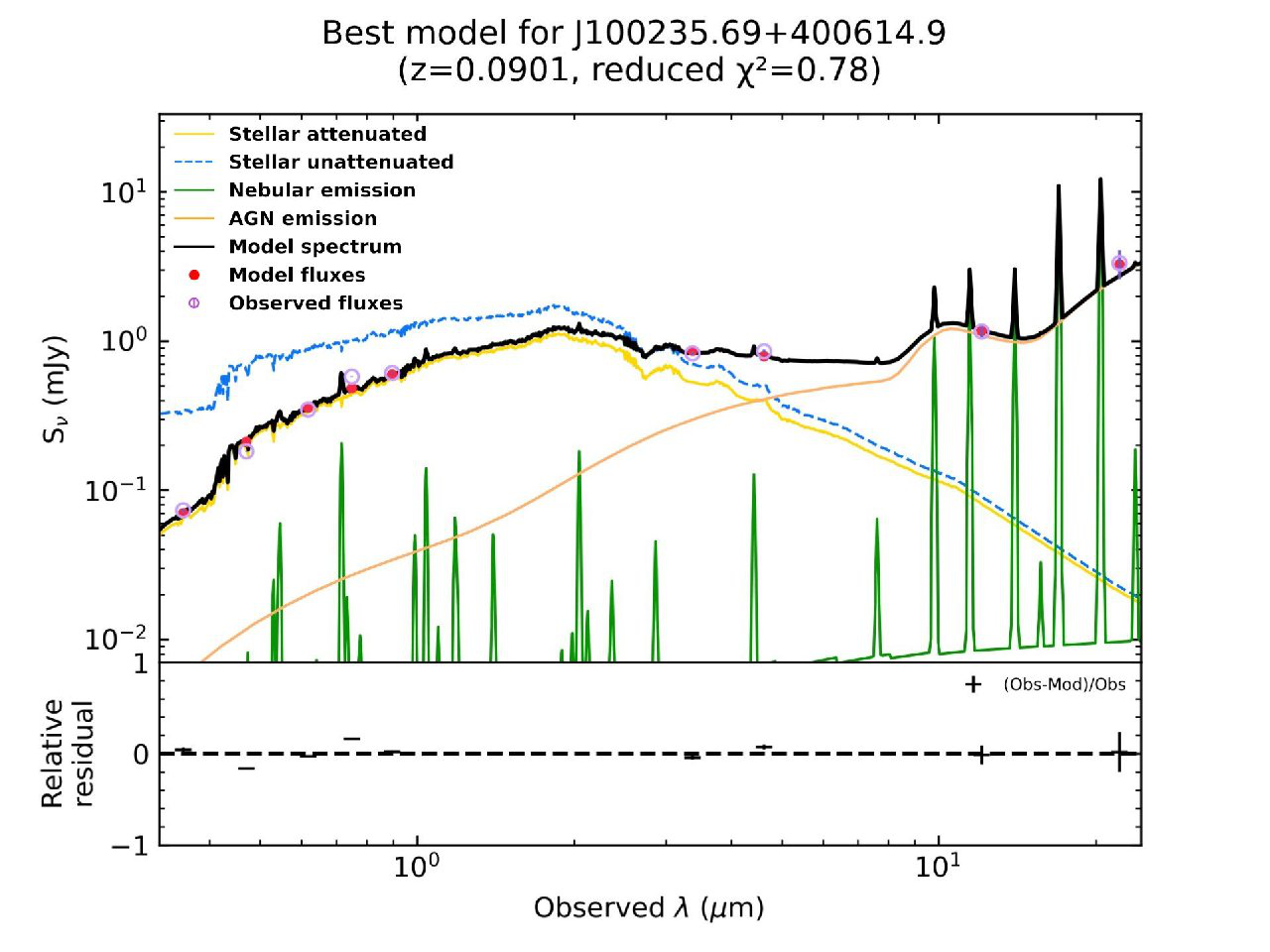}
\includegraphics[width=0.8\columnwidth]{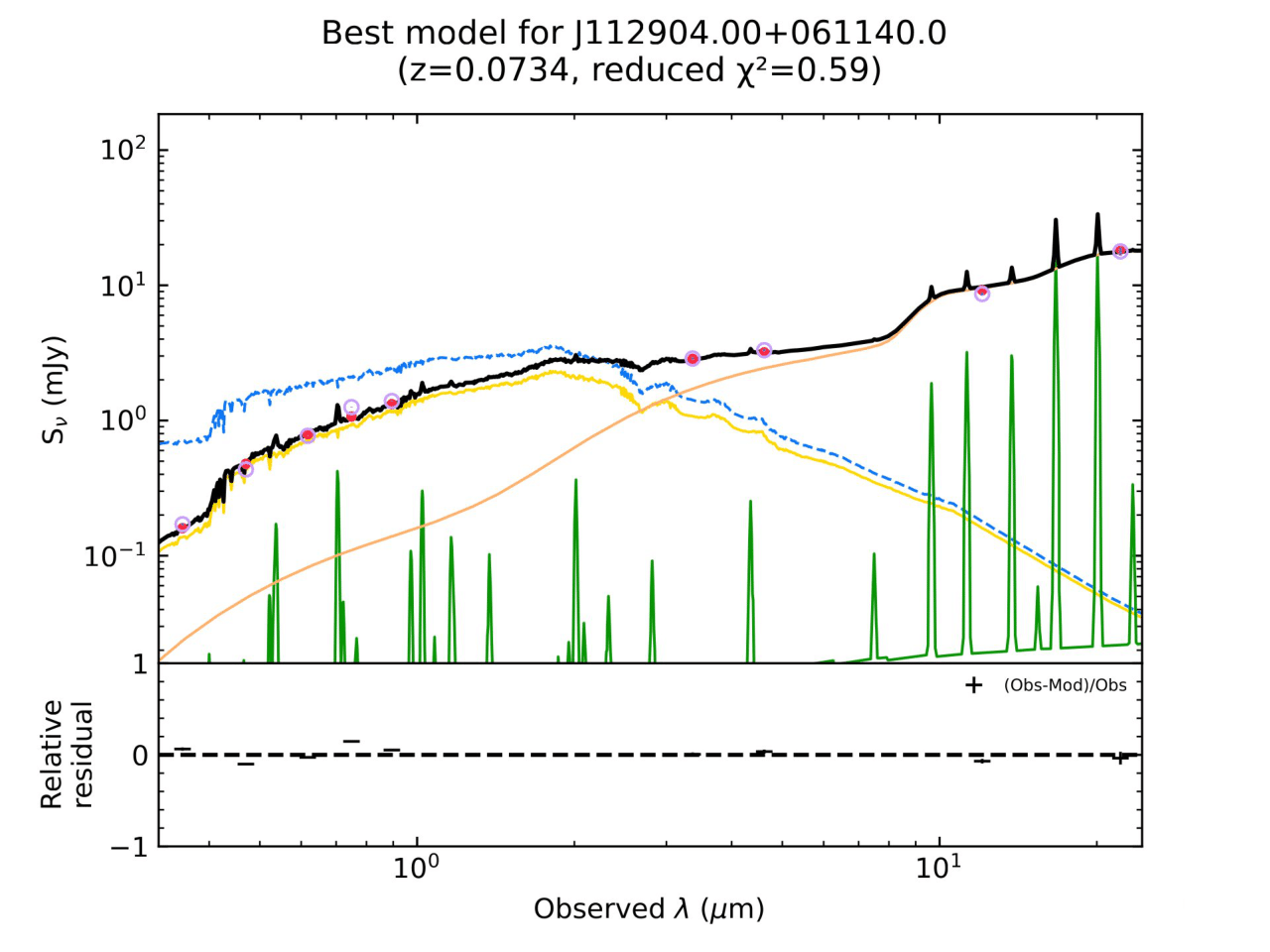}
\includegraphics[width=0.8\columnwidth]{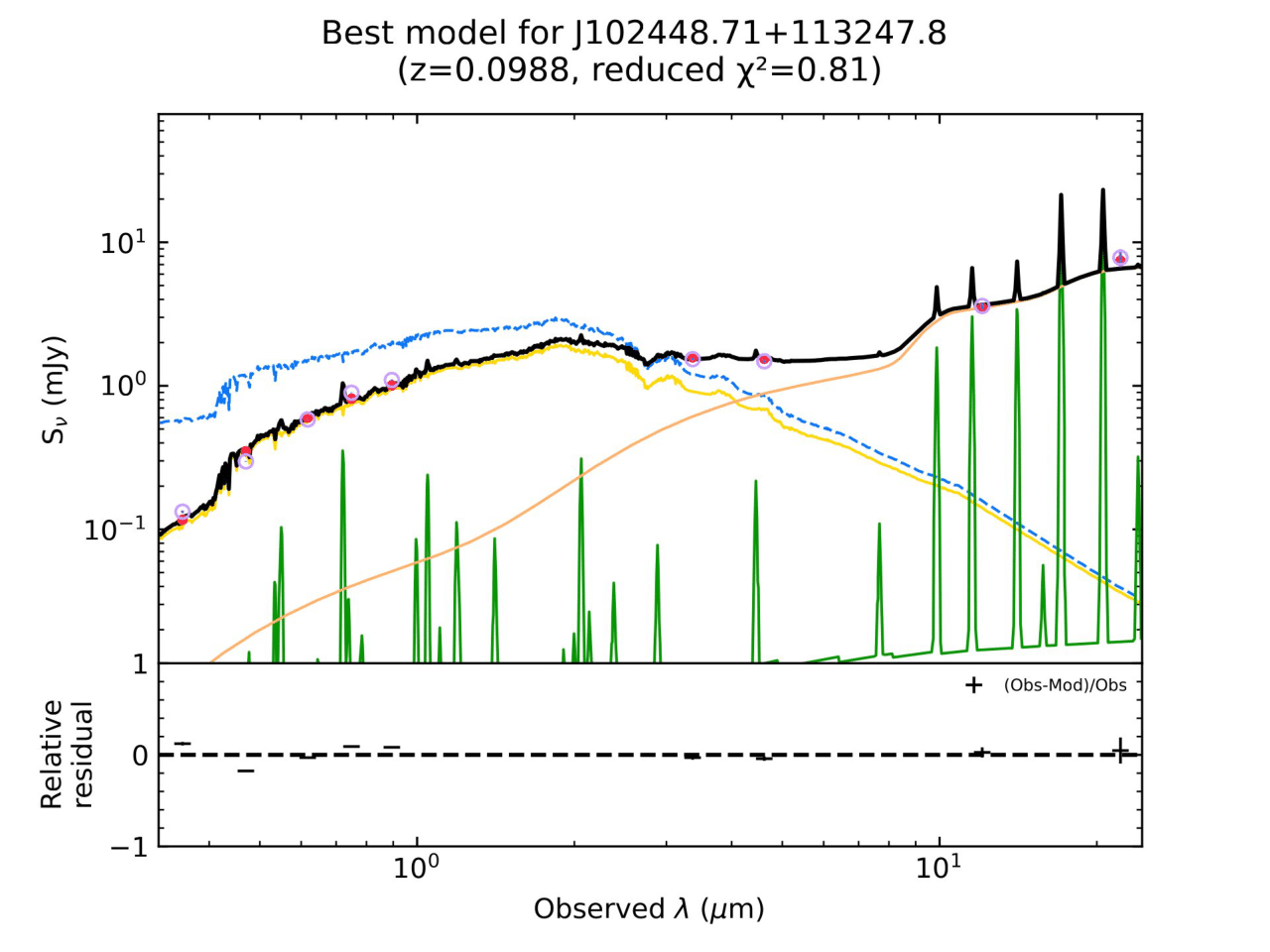}
\includegraphics[width=0.8\columnwidth]{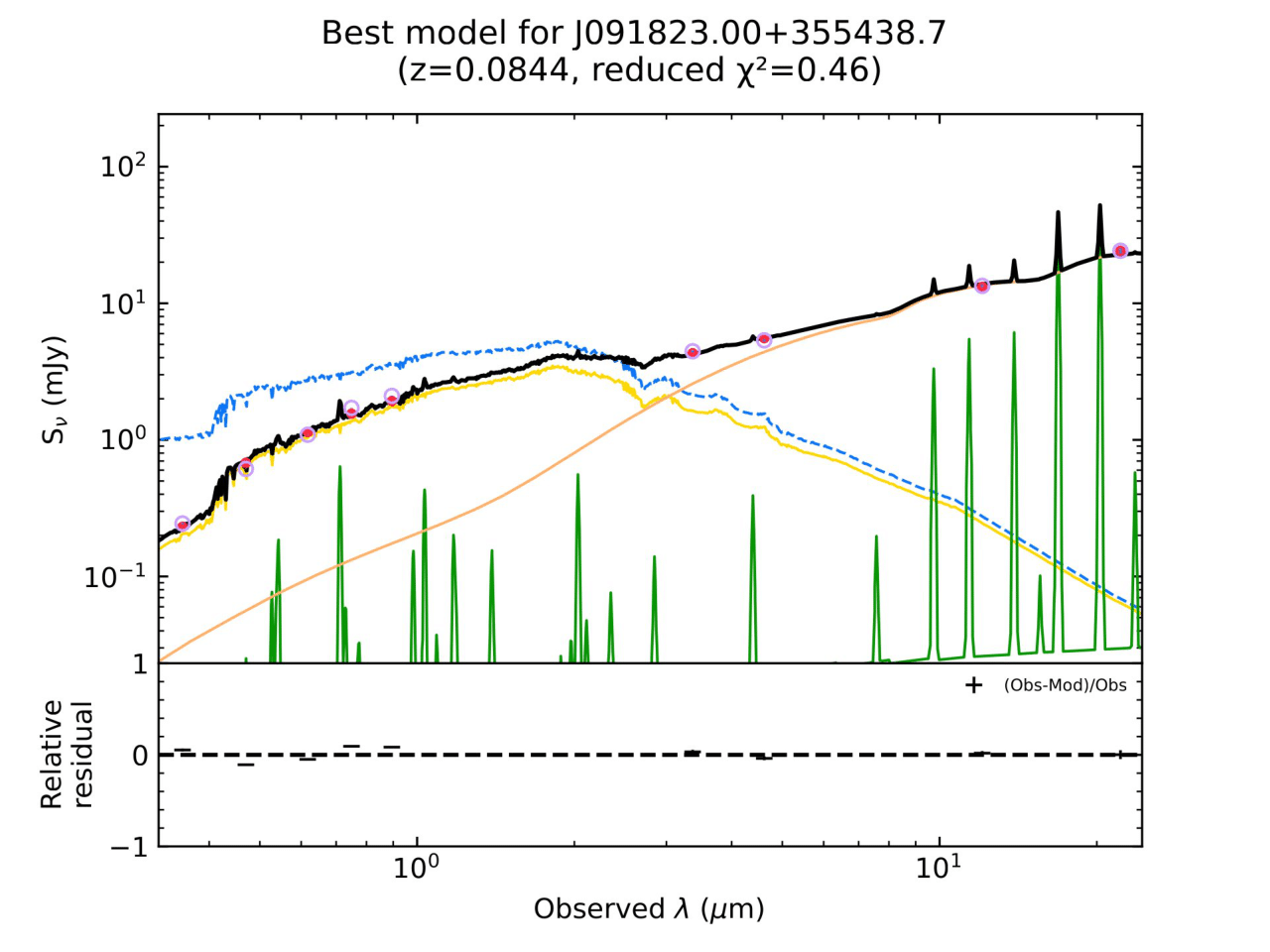}
\caption{Comparing results produced by X-CIGALE for the best SEDs in the SW sample at lower $z$ (bin 1, $\Delta 
 z = [0.05 - 0.10)$): Upper panels, left = Sy1B, right = Sy1N, lower panels, left = Sy1Bw, right = Sy1Nw.}
    \label{fig:Sy1Blowz}
\end{figure*}

\begin{figure*}
\includegraphics[width=0.8\columnwidth]{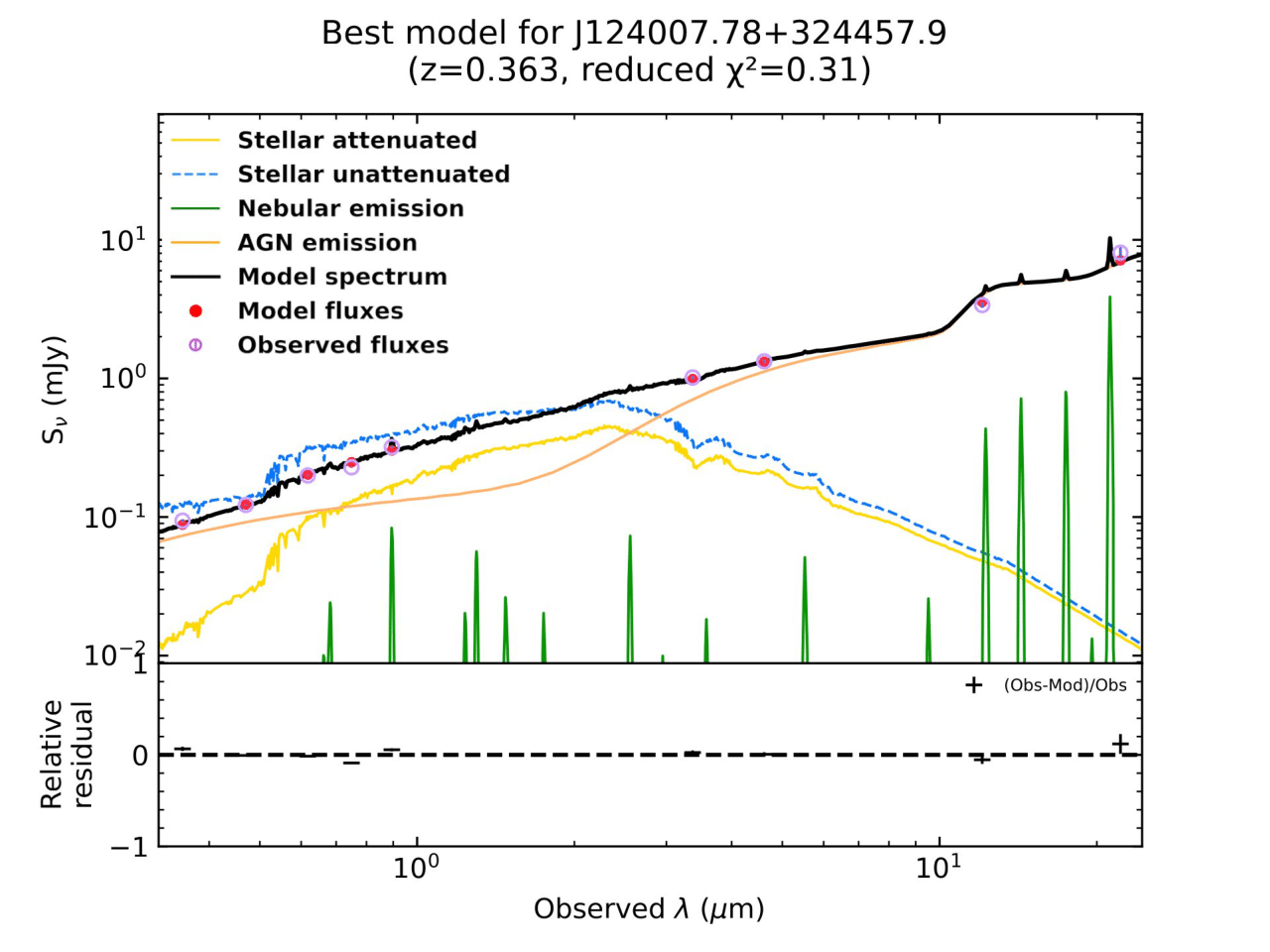}
\includegraphics[width=0.8\columnwidth]{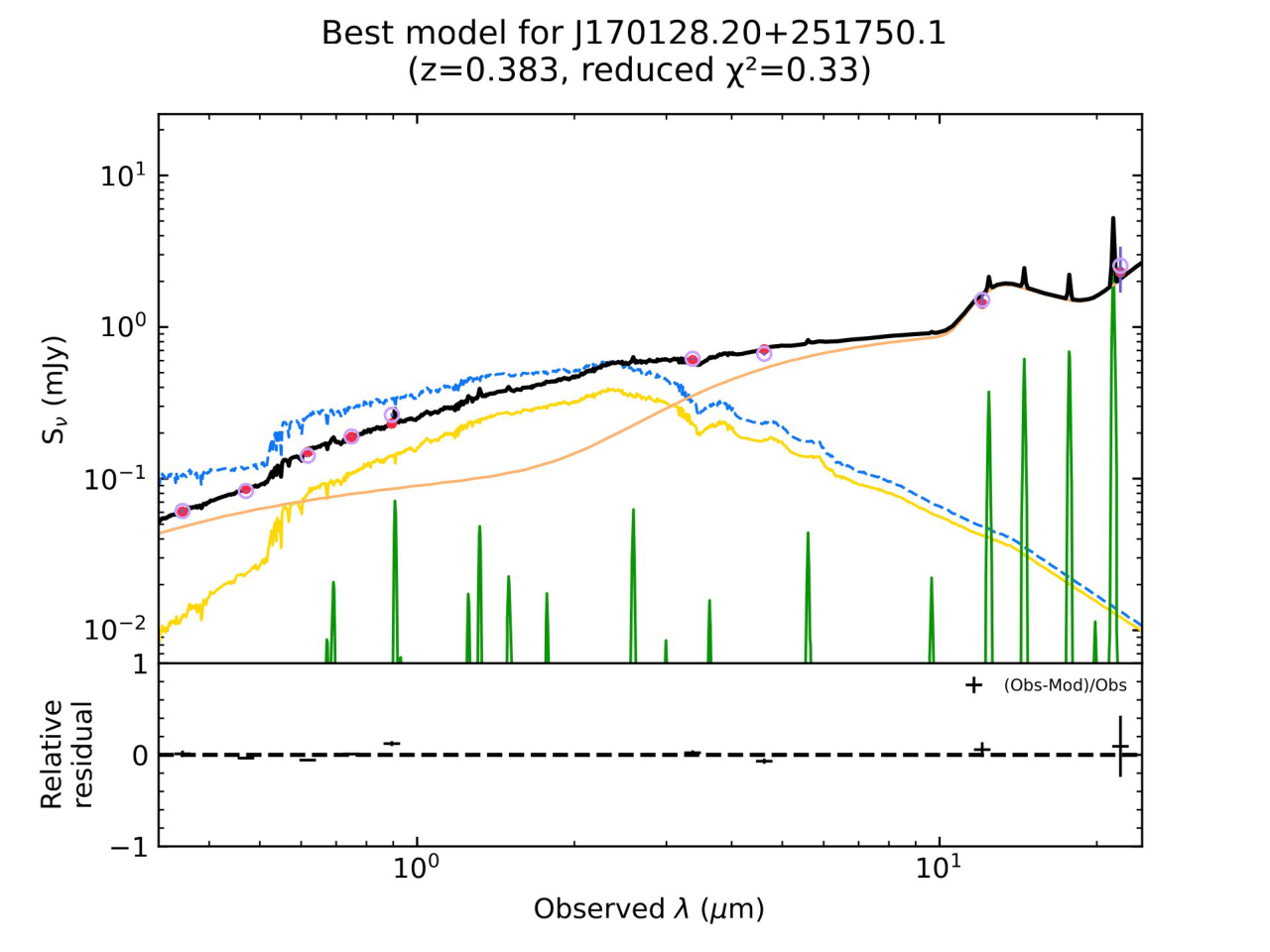}
\includegraphics[width=0.8\columnwidth]{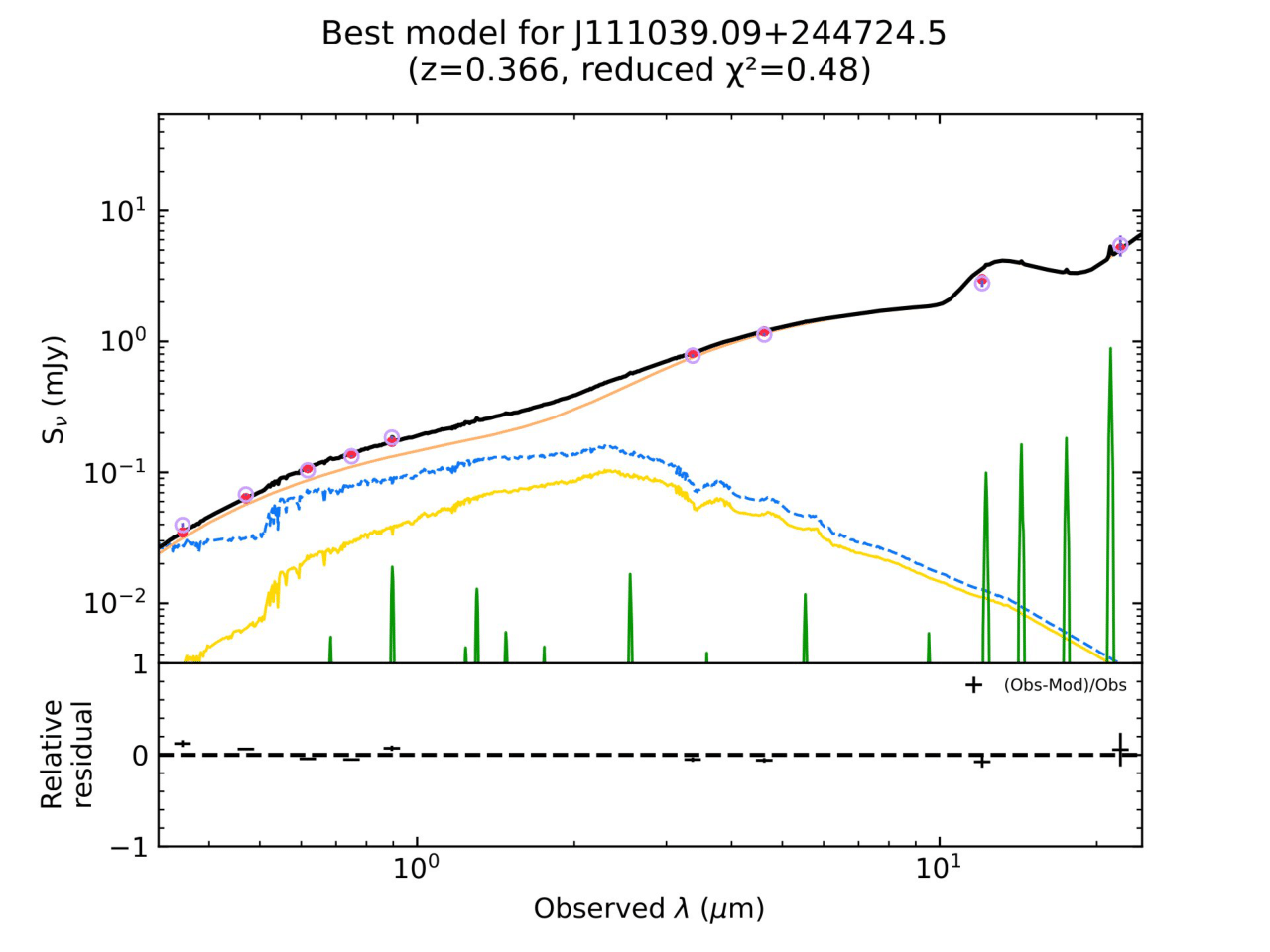}
\includegraphics[width=0.8\columnwidth]{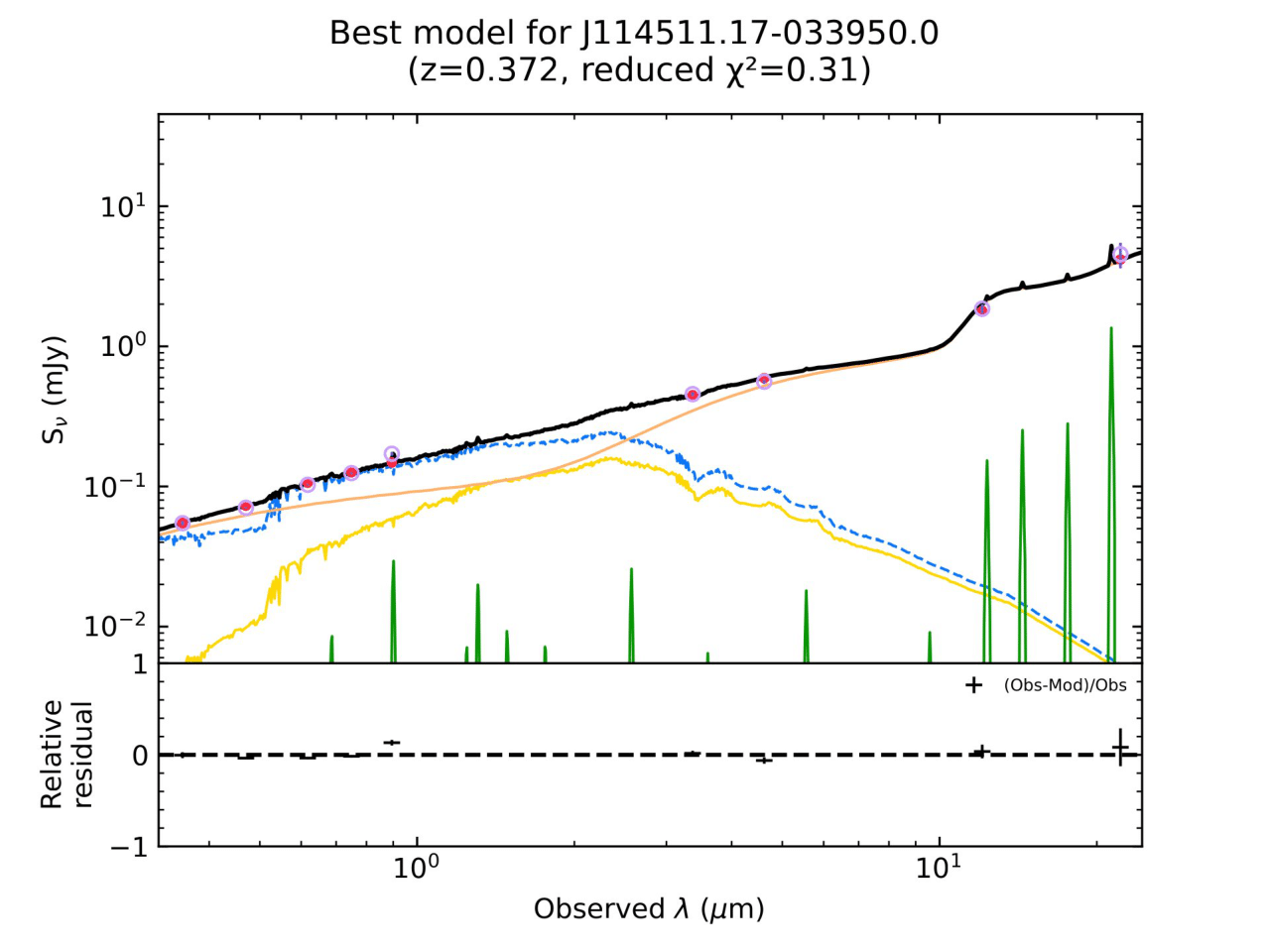}
\caption{Comparing the best Sy1 SEDs in the SW sample at higher z (bin 7, $\Delta z = [0.35 - 0.40)$): Upper panels, left = Sy1B, right = Sy1N, lower panels, left = Sy1Bw, right = Sy1Nw.}
\label{fig:Sy1_highz}
\end{figure*}

Encouraged by these results, we applied the generic SED to the ensemble of Sy1 in the SW2M sample. The medians $i$ values reported in Table~\ref{tab:SFR-fagn-i-oa_SW2M} are similar to what was obtained in the SW sample and the results of Dunns' post tests in Table~\ref{tab:dunn-SFR-fagn-i-oa_SW2M} confirm that $i$ is the same in all the spectral subgroups. The box plots for the $\chi^{2}$ with outliers are shown in the bottom panel of Figure \ref{fig:Sy1-chi}. The outlier percentages are 10.5\% in the Sy1B, 10\% in the Sy1Bw, 5.6\% in the Sy1N, and 7.8\% in the Sy1Nw. Comparing in Table~\ref{tab:chiSy1} the medians for the $\chi^{2}$ in the SW2M and SW samples, a weak trend for a slightly higher values is detected in the SW2M sample, which suggests that adding MIR photometric data might introduce some slight differences in the parameters of the generic SED relative to the torus ($f_{AGN}$, $oa$ and SFR). Consequently, we will present the results separately in the next section. 

\section{Results} 
\label{Results}

\begin{figure*}
\includegraphics[width=0.85\columnwidth]{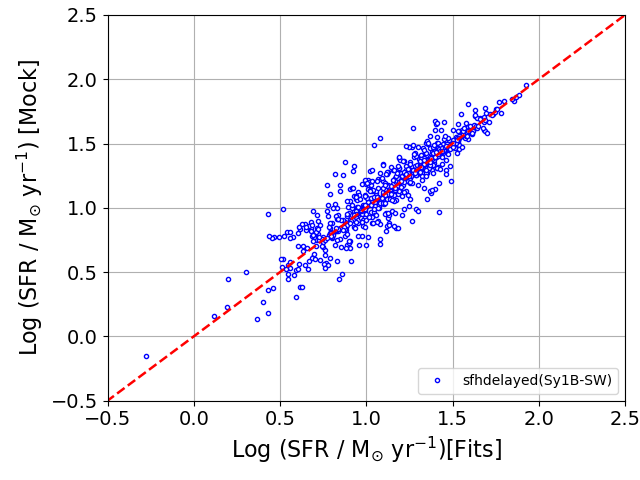}
\includegraphics[width=0.85\columnwidth]{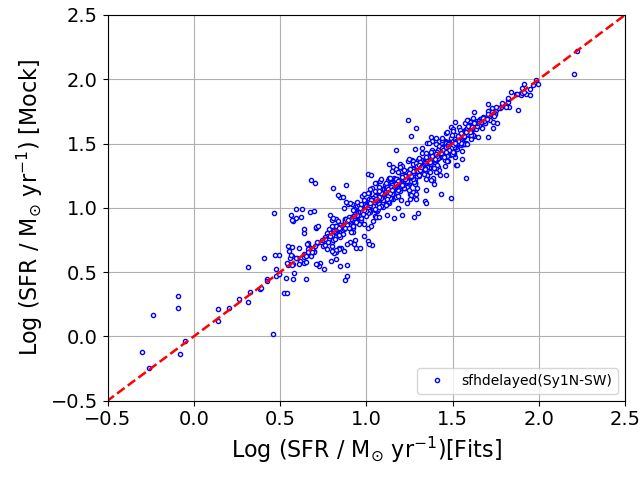}
\includegraphics[width=0.85\columnwidth]{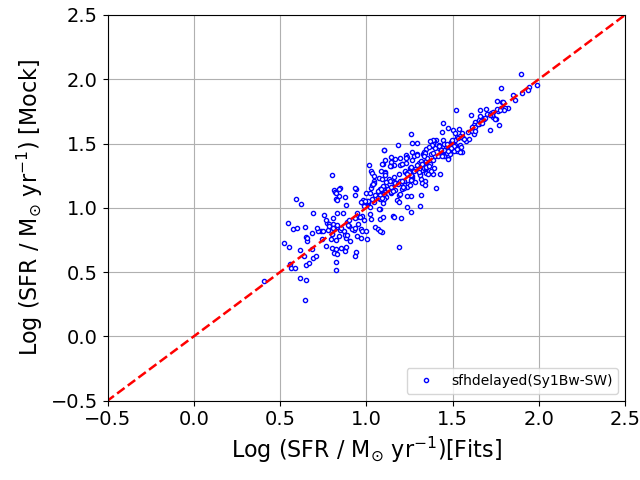}
\includegraphics[width=0.85\columnwidth]{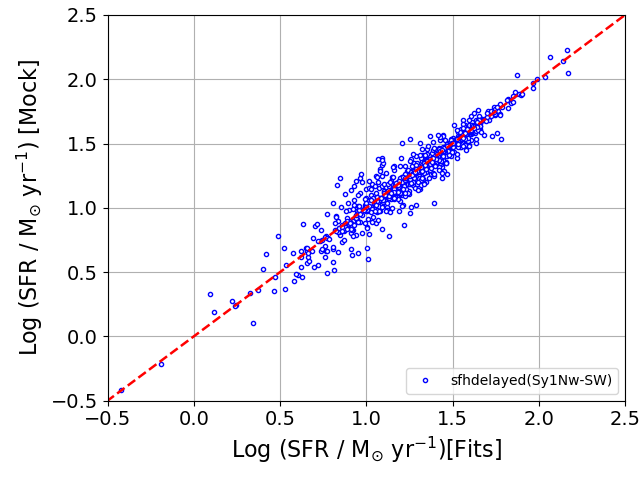}
\caption{Comparing the SFRs from the best fit SEDs with the SFRs from the mock catalogues; the red dash lines are linear regressions, with slopes close to one and coefficient of correlation of 0.92 for the Sy1B and 0.96 for the Sy1N. }
    \label{fig:Sy1B-mock}
\end{figure*}

In Figure~\ref{fig:Sy1Blowz} 
we compare the SEDs obtained using the Sy1 generic SED model on the galaxies in the sample SW at lower z. Each SED is composed of four components: 1- AGN (orange line), 2- unattenuated stellar populations (blue line), 3- stellar populations attenuated by dust extinction (yellow line), and 4- nebular emission lines (green line). The sum of the components is the continuous black line. Also shown are the fluxes from the model (red dots) and the observed fluxes (open circles). The goodness of the fits can be appreciated directly by visual inspection of the SEDs and from the residuals traced below each of them.
The most intriguing feature of theses SEDs 
is that, in any spectral subgroup, with or without AGN wind, the AGN component is only dominant in the IR. According to X-CIGALE, even after attenuation by dust, the stellar populations formed in majority of massive young stars---the blue component having the flat continuum shape typical of active star forming regions---dominates the UV-optical range of the SED. Consequently, star formation in these galaxies is very active, in fact, much more active than previously estimated by \citet[][]{2020Torres-Papaqui}. 

In Figure~\ref{fig:Sy1B-mock}, we compare the bets fit SFRs for the whole sample with the reciprocal SFRs produced by the mock catalogues. The option to produce a mock catalogue in X-CIGALE is used to verify the robustness of the best fits: using the output values as initial inputs, the program repeat the fitting process varying the parameters, allowing to estimate wether the solutions converge or not to the best fit values. Comparing the SFRs in  Figure~\ref{fig:Sy1B-mock}, one-to-one relations are obtained showing that the solutions in all the spectral subgroups converge. In Figure~\ref{fig:Sy1B-mock}, the red-dash lines are linear regressions with slopes close to one and coefficients of correlation 0.92 for the Sy1B/Sy1Bw and 0.96 for the Sy1N/Sy1Nw, implying that the SFRs produced by X-CIGALE are robust. This result supports the conclusion that in all these spiral galaxies, independently of their spectral subgroup and on the presence or not of AGN wind, star formation is very active. 

\begin{table}
\centering
\caption{Median values produced by the Sy1 generic SED on galaxies in the sample SW.
}
\label{tab:SFR-fagn-i-oa} 
\begin{tabular}{|c|c|c|c|c|}
\hline
Subgroup & $\log (\rm{SFR})$ & $f_{\rm AGN}$ & \textit{i}  &\textit{oa}   \\ 
&(M$_\odot$ yr$^{-1})$&&(deg.)&(deg.)\\
\hline
Sy1B & 1.17 & 0.44 & 9.76 & 39.72 \\ 
Sy1Bw & 1.21 & 0.51 & 9.66  & 40.71 \\ 
 Sy1N & 1.25 & 0.20 & 9.78 & 35.19 \\ 
Sy1Nw & 1.27 & 0.30 & 9.72 & 39.96 \\ \hline
\end{tabular}               

\end{table}

\begin{figure}
	\includegraphics[width=\columnwidth]{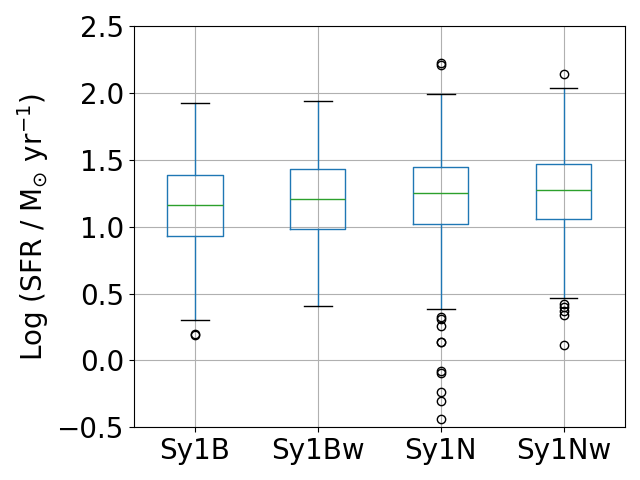}
    \caption{Box-whisker plots for the SFRs produced by the Sy1 generic SED in the SW sample.}
    \label{fig:Sy1FR_BxPl}
\end{figure}

\begin{table}
\centering
\caption{Summaries of Dunn’s Multiple Comparisons Test for spectral subgroups in the sample SW. Levels of significance as explained in Table 2.}
\label{tab:dunn-SFR-fagn-i-oa}
\begin{tabular}{|c|c|c|c|c|}
\hline
Pairs & SFR & $f_\textit{AGN}$ & \textit{i}  &\textit{oa}   \\ 
\hline
Sy1B-Sy1Bw  & * & ** & ns & *** \\
Sy1B-Sy1N   & **** & **** & ns  & * \\ 
Sy1B-Sy1Nw  & **** & **** & ns & * \\ 
Sy1Bw-Sy1N  & ns & **** & ns & **** \\ 
Sy1Bw-Sy1Nw &** &****& ns & ns \\ 
Sy1N-Sy1Nw  & * &**** & ns & **** \\ 
\hline
\end{tabular}
\end{table}

How intense is the star formation can be better estimated by tracing the box-plots for the SFRs in the four spectral subgroups. This is done in Figure~\ref{fig:Sy1FR_BxPl}, where, judged from the first and third quartiles, the SFRs are shown to vary between 10 and 32 M$_\odot$ yr$^{-1}$. This is compatible either with normal star formation in late-type spirals or in starburst galaxies \citep{2012Kennicutt}. However, since the host galaxies of the Sy1 tend to be early-type spirals \citep{2020Torres-Papaqui}, the starburst interpretation might be closer to reality. In Figure~\ref{fig:Sy1FR_BxPl} we also detect a trend for SFR in Sy1N and Sy1Nw to be higher than in Sy1B and SyBw. This trend also appeared in Table~\ref{tab:SFR-fagn-i-oa} comparing the medians. Note that all the statistics were calculated after eliminating the outliers identified in Figure~\ref{fig:Sy1-chi}. To verified the statistical significance of these differences, we apply a Kruskal–Wallis test, which detect significant differences at a level of confidence of 95\%. In Table~\ref{tab:dunn-SFR-fagn-i-oa} the results for the Dunn's post-test confirms at the highest level of significance that Sy1N have higher SFR than Sy1B. This difference also applies comparing Sy1Nw with Sy1Bw, while the significance is lower comparing Sy1 with and without AGN wind within the same spectral subgroup. Consequently, we distinguish two origins for the differences in SFR, the most important being a difference related to the spectral group, SFR being higher in N than B subgroup, then Sy1 with AGN wind having slightly higher SFR than Sy1 without wind.  

\begin{figure}
\includegraphics[width=\columnwidth]{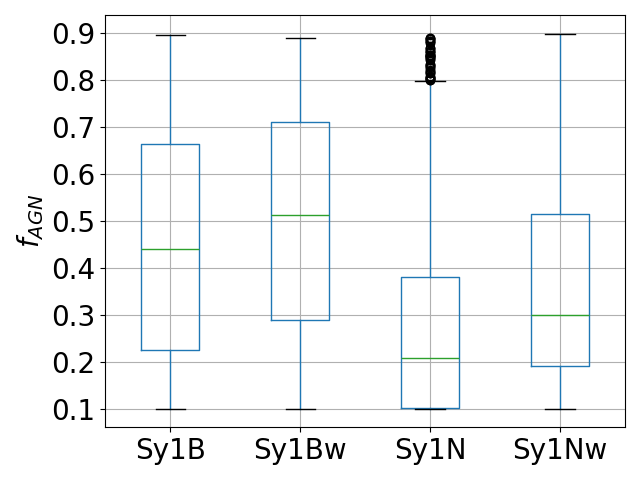}
\caption{Box-whisker plots for the fraction of AGN, $f_{AGN}$, as produced by the Sy1 generic model for the galaxies in the SW sample.}
\label{Sy1fAGN_BxPl}
\end{figure}

\begin{figure}
\includegraphics[width=\columnwidth]{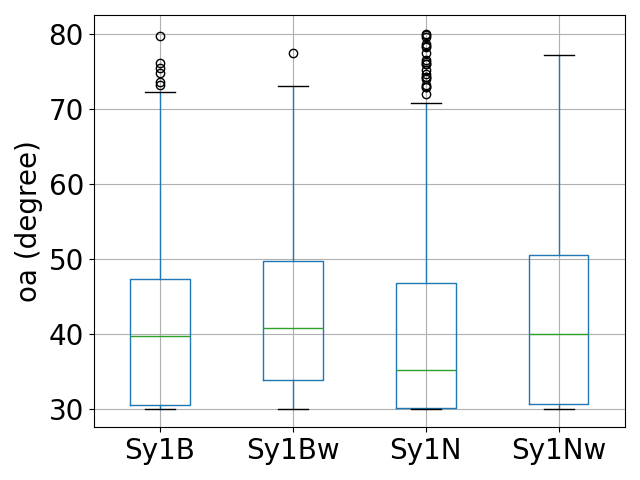}
\caption{Box-whisker plots for the opening angle, \textit{oa}, as produced by the Sy1 generic model for the galaxies in the SW sample.}
\label{fig:Sy1oa_BxPl}
\end{figure}

In Figure~\ref{fig:Sy1_highz} we compare the best fits as generated by the Sy1 generic SED at higher $z$ (bin 7 in Table~\ref{tab:BIN_SW}). The most obvious difference in the SED appears to be an increase of the AGN component in UV/Opt, the AGN continuum dominating the attenuated SF component, in part in the Sy1 without wind and completely in Sy1 with wind. 

In Figure~\ref{Sy1fAGN_BxPl}, we compare the box-whisker plots for the fraction of AGN, $f_\textit{AGN}$, which in X-CIGALE is calculated over the whole SED. There are obvious differences between the spectral subgroups, clearly appearing also in the medians reported in Table~\ref{tab:SFR-fagn-i-oa}: Sy1B have higher $f_\textit{AGN}$ than Sy1N and Sy1 with wind have higher $f_\textit{AGN}$ than Sy1 without wind. The results of Dunn's post-test reported in Table~\ref{tab:dunn-SFR-fagn-i-oa} confirm the differences between all the pairs.

Surely the most consequential point about the differences in $f_{AGN}$ is that they are similar to those obtained by \citet{2020Torres-Papaqui} for the differences in AGN luminosities, L$_{AGN}$. This implies that there is a direct connection between $f_\textit{AGN}$ in the generic SED of the Sy1 as produced by X-CIGALE and L$_{AGN}$ as deduced directly from their SDSS spectra. This is one important argument in favor of the generic SED model, implying that this SED (although based on a rough approximation of the torus structure) can reproduce an important physical characteristics of AGN related to the accretion rate of matter by the SMBH.  

\begin{figure*}
\includegraphics[width=0.8\columnwidth]{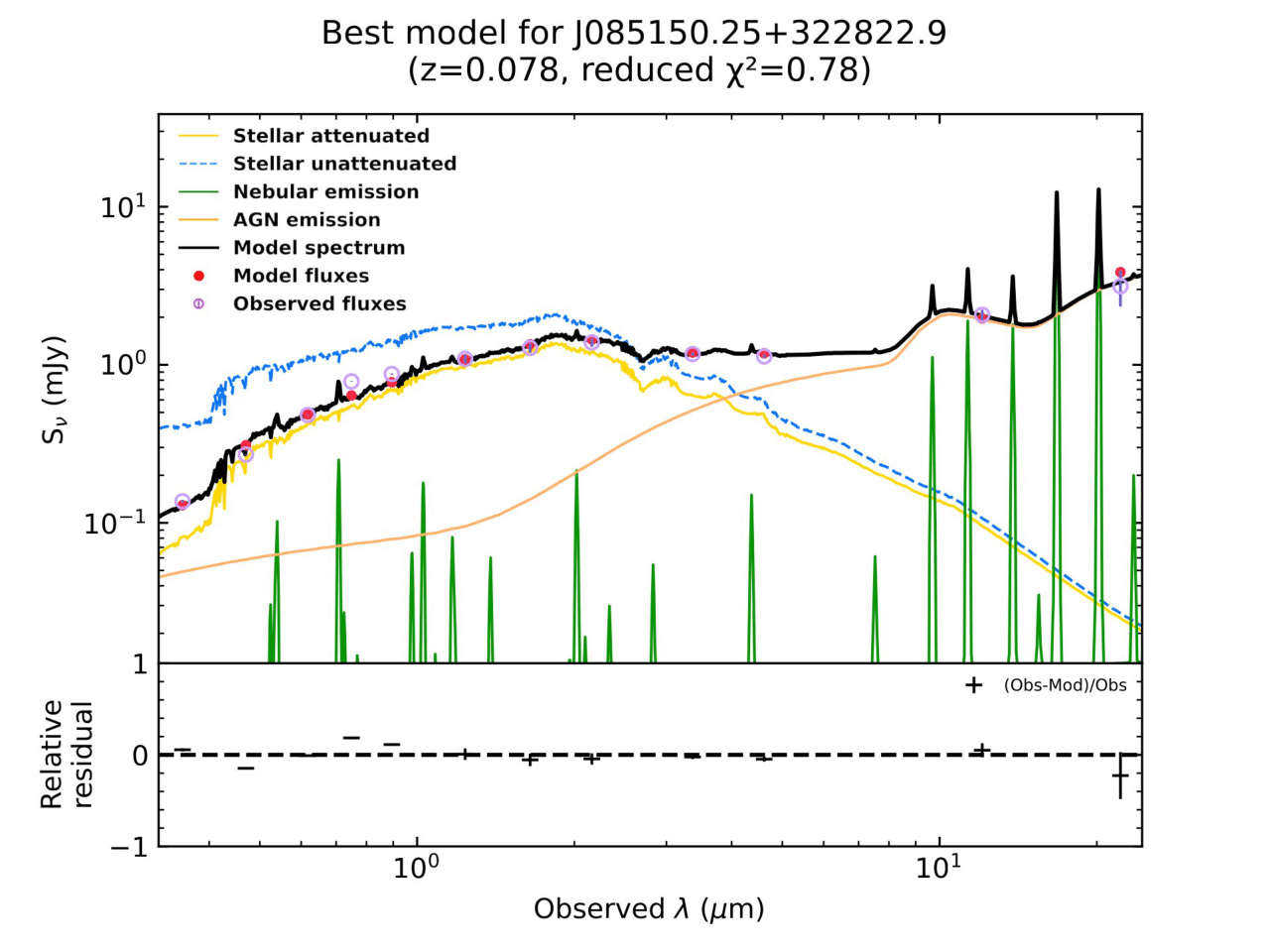}
\includegraphics[width=0.8\columnwidth]{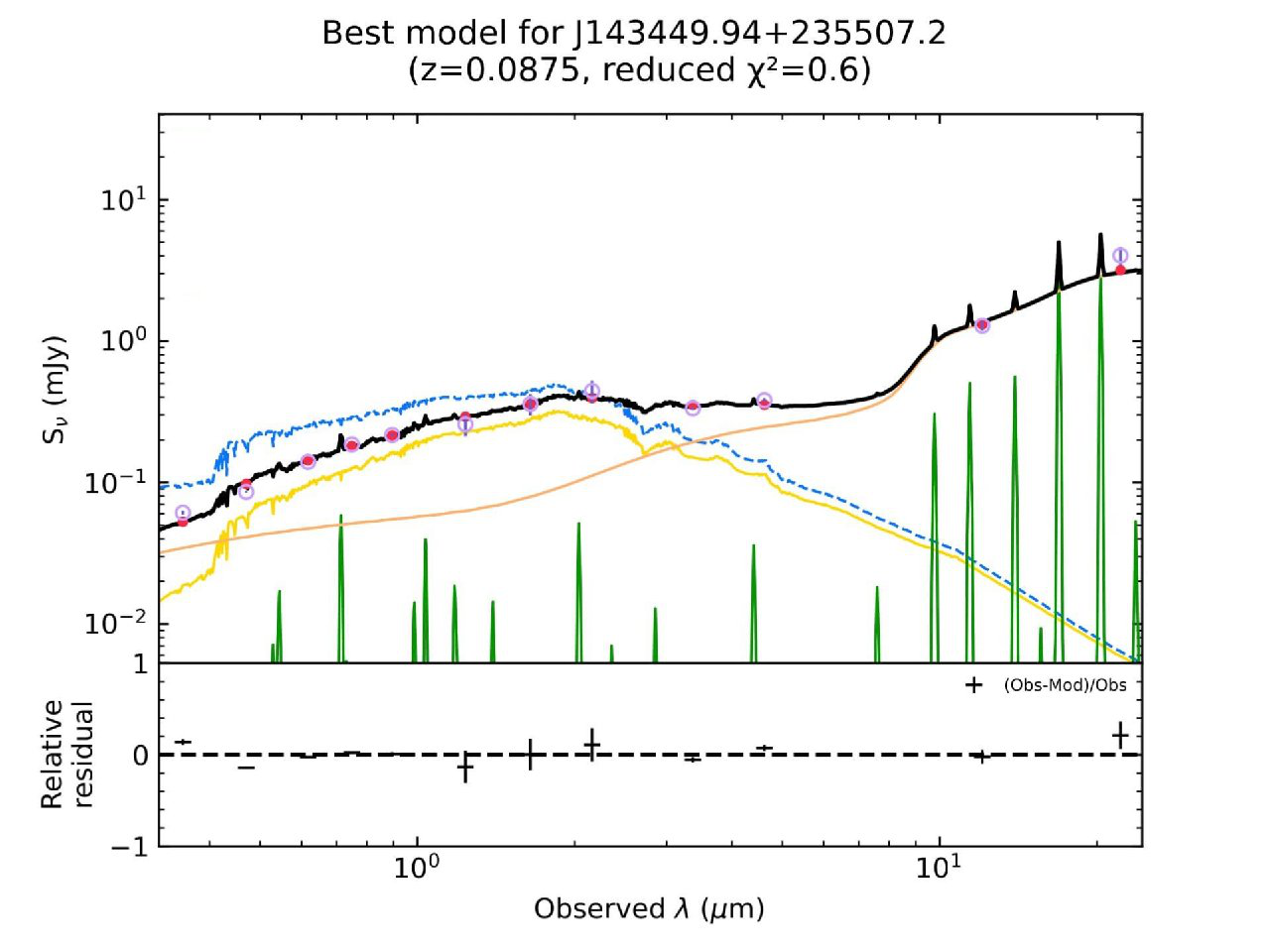}
\includegraphics[width=0.8\columnwidth]{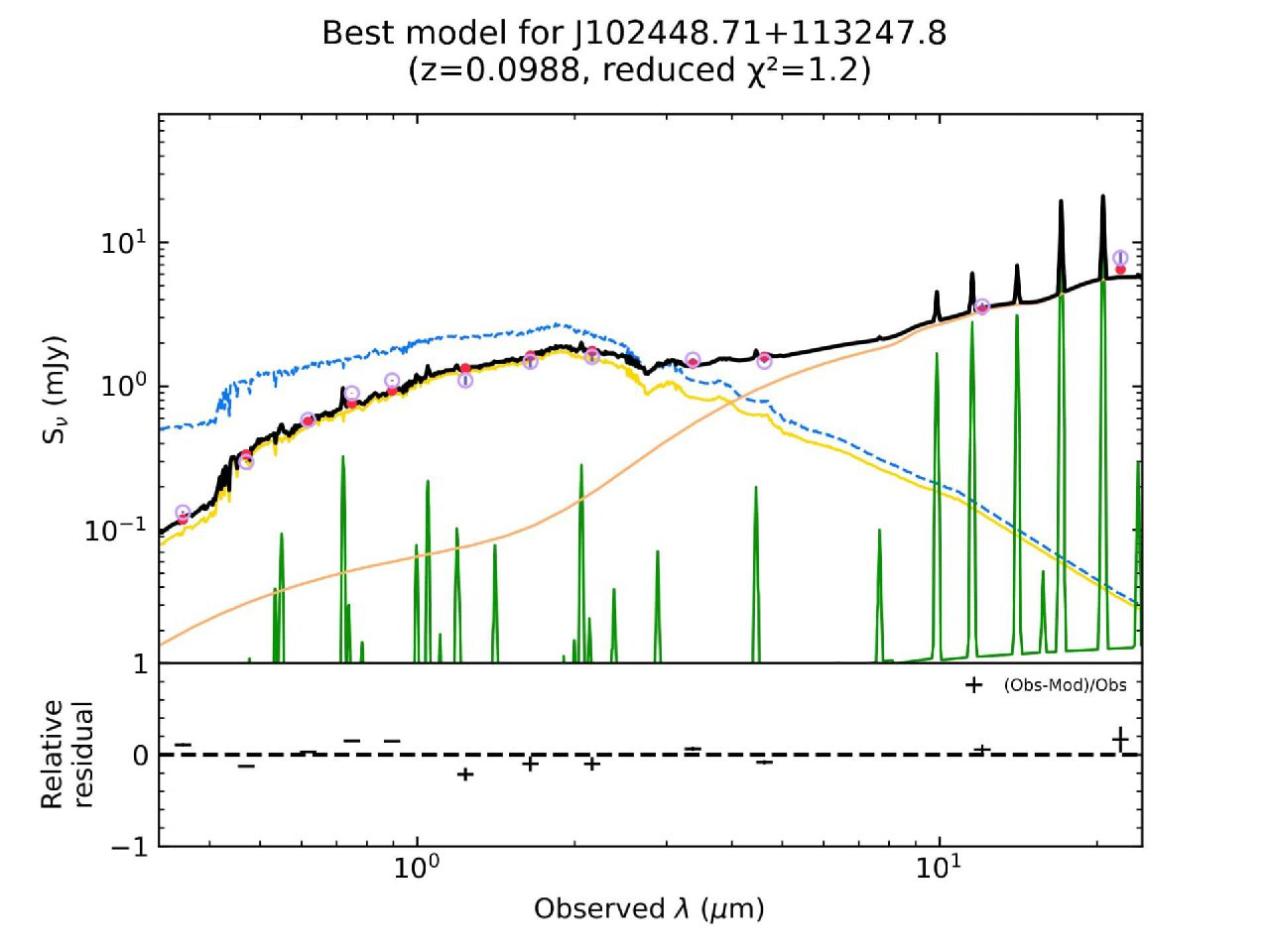}
\includegraphics[width=0.8\columnwidth]{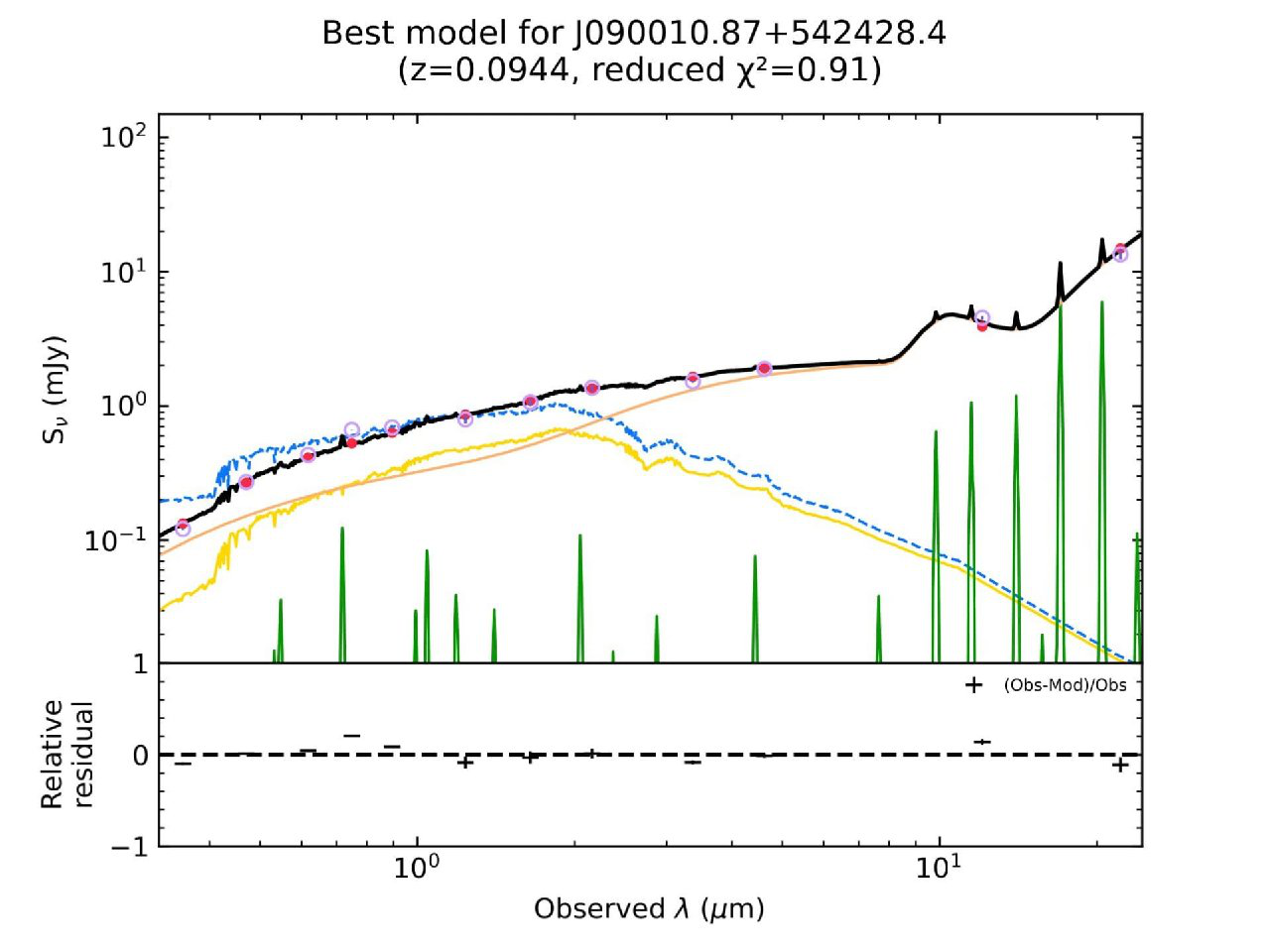}
\caption{Comparing results produced by X-CIGALE for the best SEDs in the SW2M sample at lower $z$ (bin 1, $\Delta 
 z = [0.05 - 0.10)$): Upper panels, left = Sy1B, right = Sy1N, lower panels, left = Sy1Bw, right = Sy1Nw.}
    \label{fig:Sy1-SW2M_lowz}
\end{figure*}

\begin{figure*}
\includegraphics[width=0.8\columnwidth]{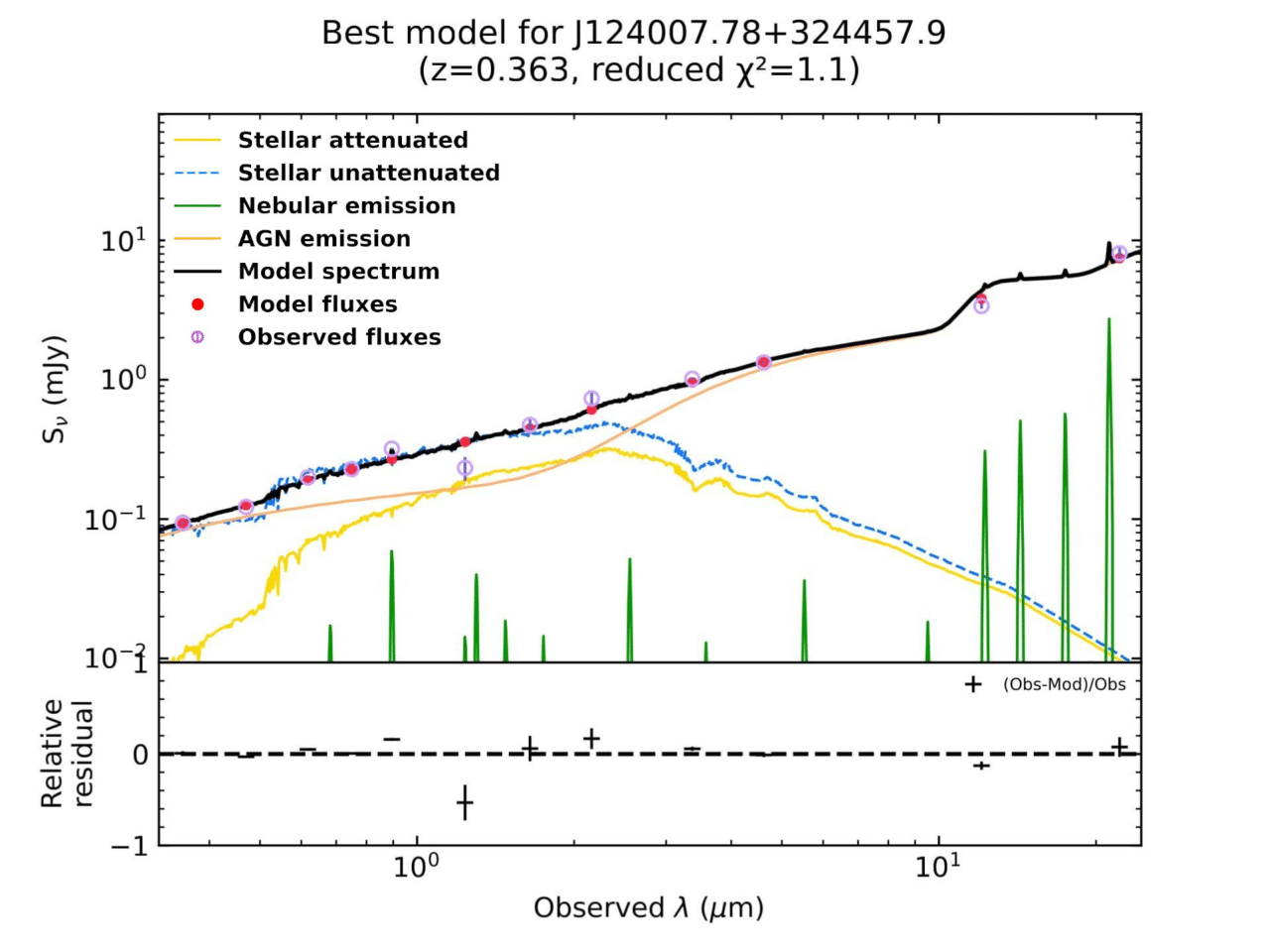}
\includegraphics[width=0.8\columnwidth]{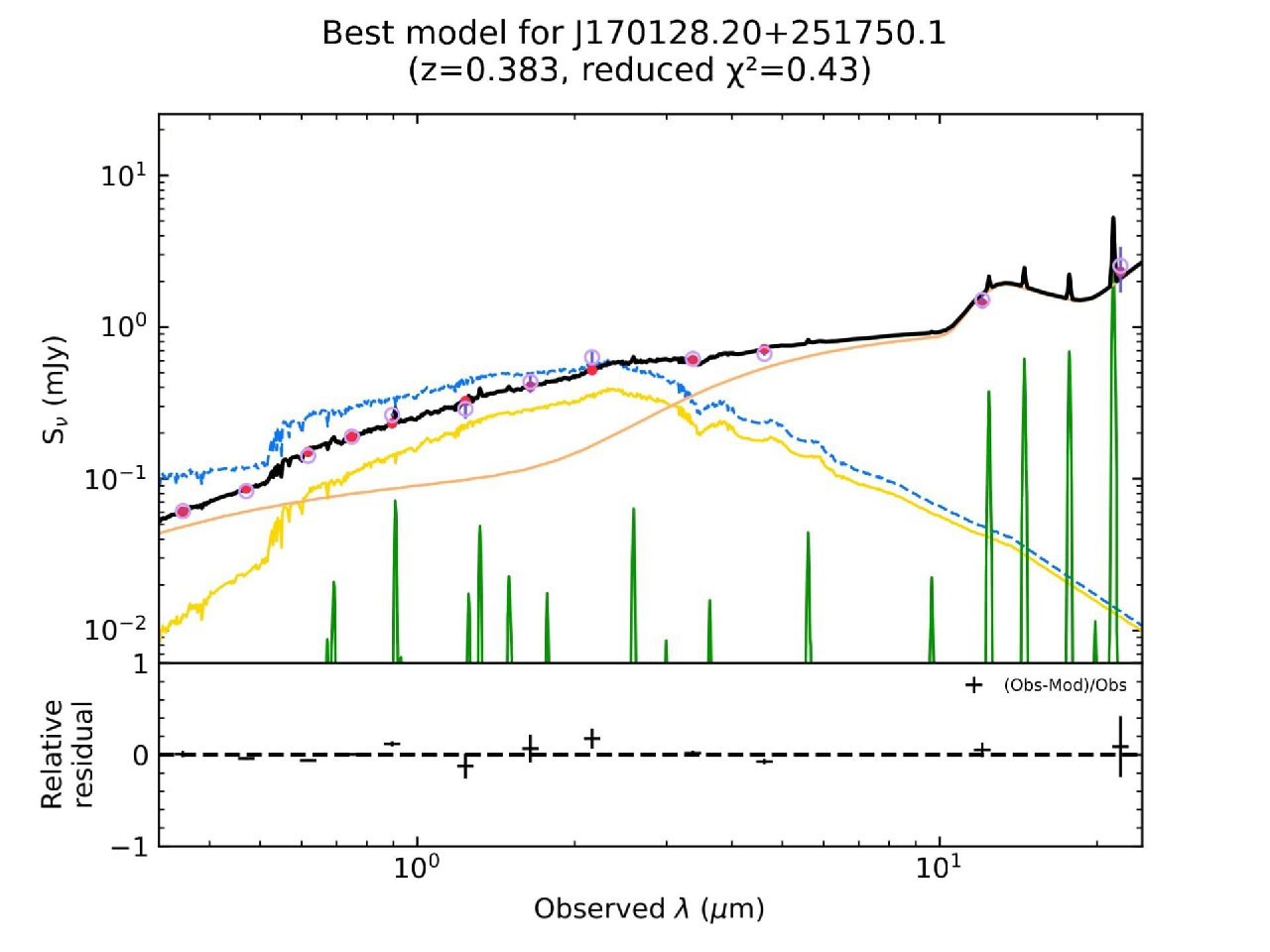}
\includegraphics[width=0.8\columnwidth]{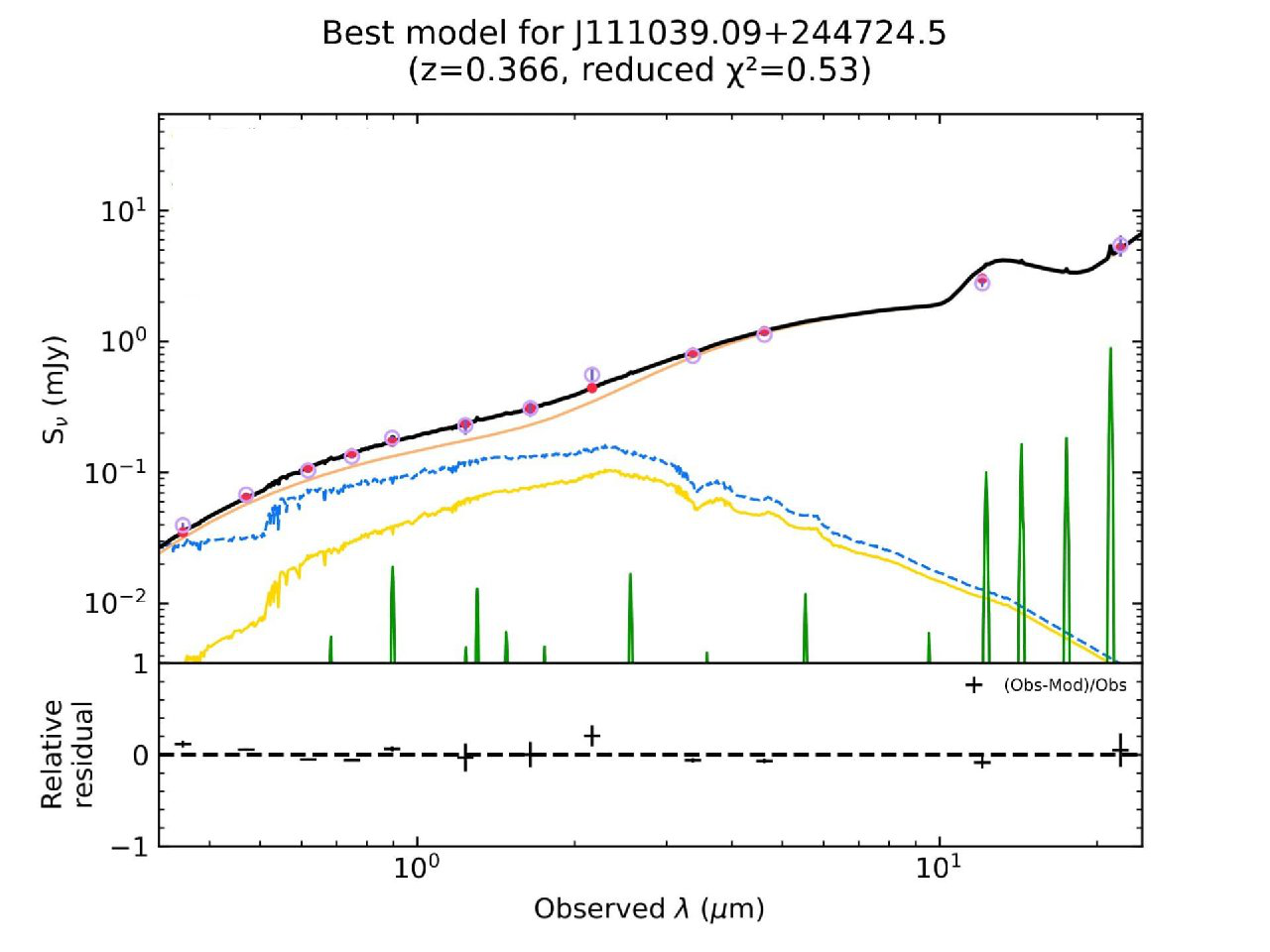}
\includegraphics[width=0.8\columnwidth]{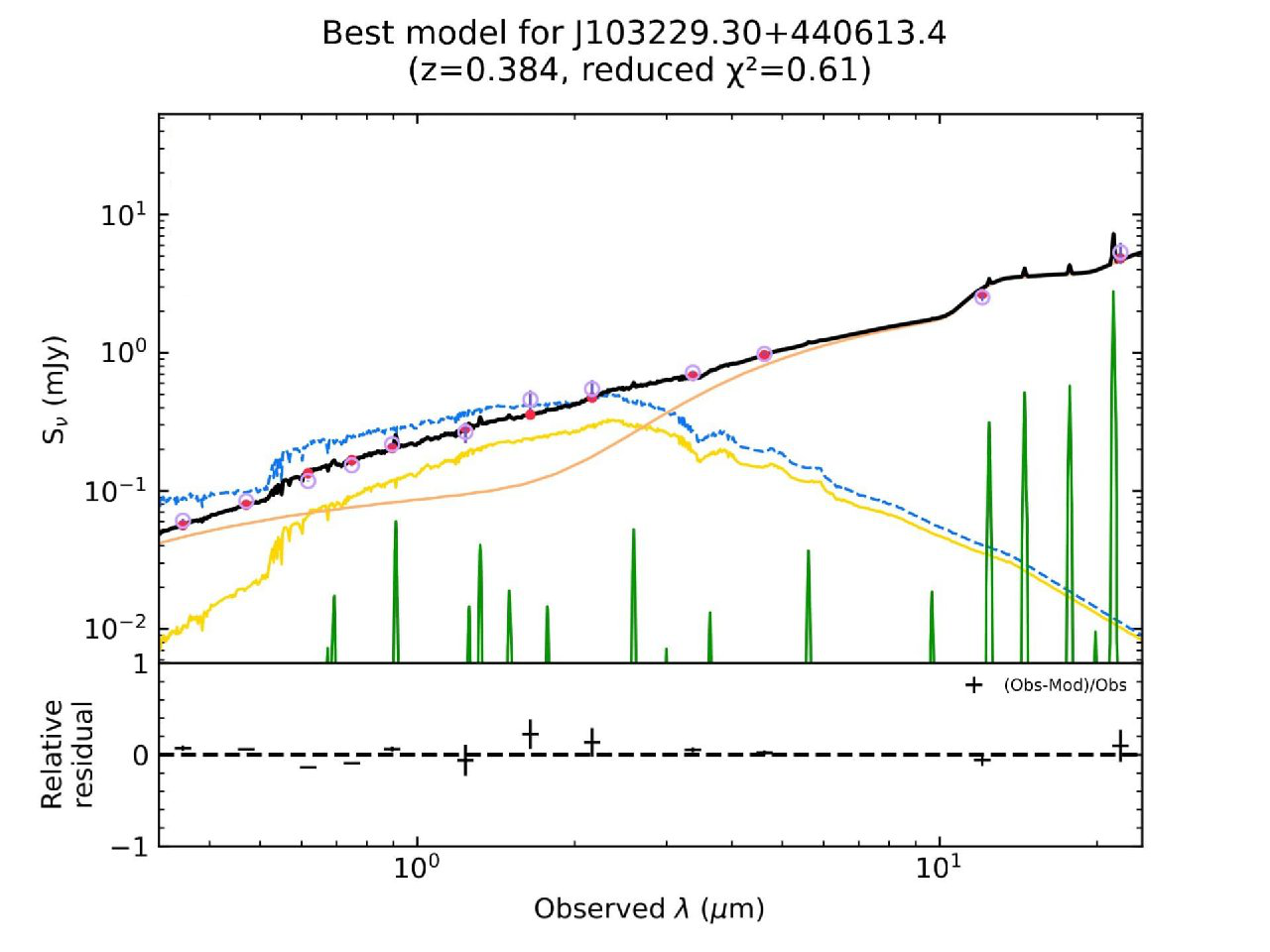}
\caption{Comparing the best Sy1 SEDs in the SW2M sample at higher z (bin 7, $\Delta z = [0.35 - 0.40)$): Upper panels, left = Sy1B, right = Sy1N, lower panels, left = Sy1Bw, right = Sy1Nw.}
\label{fig:Sy1-SW2M_highz}
\end{figure*}

As we explained in our introduction, another advantage of using X-CIGALE is that we should get some information about the torus itself and, more specifically, how it obstructs the central engine. In particular, one of our initial hypotheses to explain the Sy1N spectral subgroup was that narrow lines appear in these AGN because we see them at intermediate LOS angles relative to the tori. However, since \textit{i} is small in the generic model, and does not seem to vary in the spectral subgroups, this hypothesis seems now definitely falsified. Therefore, another characteristic must be involved, either of the torus or of the broad line region (BLR) filling matter to the accretion disk. In the generic model, the second parameters that we allowed to vary was the half-opening angle, \textit{oa}, which yields information about the covering factor of the torus, $f_{cov}$, that is, the part that is illuminated by the AGN. Assuming the covering factor of the torus is the same as the covering factor of the BLRs \citep{2009Gaskell}, one can thus get information on both structures at the same time.  

In Figure~\ref{fig:Sy1oa_BxPl} we compare the box-whisker plots of \textit{oa} in the different spectral subgroups and report the medians in Table~\ref{tab:SFR-fagn-i-oa}. It seems clear that there is a trend for Sy1B to have larger \textit{oa} than Sy1N and for Sy1 with wind to have larger \textit{oa} than their counterparts without wind (no difference appear between Sy1Nw and Sy1Bw). In terms of the torus, the difference between Sy1 with and without wind seems obvious. Referring to Figure~6 in \citet{2009Gaskell}, when a wind is present, X-CIGALE might interpret the wind as part of the torus, explaining why the covering factor \textit{oa} increases. However, in the case of the differences between the Sy1N and Sy1B, the interpretation in terms of the torus is less clear.

\subsection{Results for the Sy1 in the SW2M sample} \label{SW2M}

Applying the generic SED to the galaxies in the sample SW2M, we compare the best fits in Figure~\ref{fig:Sy1-SW2M_lowz} and Figure~\ref{fig:Sy1-SW2M_highz} at lower and higher $z$ respectively. Note that based on the residuals the three fluxes from 2MASS are relatively well fitted in all the best SEDs. In general, we observe the same behaviour than what we observed before for the Sy1 in the SW sample: at low $z$, the AGN component is not  dominant in the UV/Opt, except maybe in the Sy1Nw, then becoming fully dominant at high $z$ in the Sy1B/Sy1Bw but only partially dominant in the Sy1N/Sy1Nw. 

\begin{figure*}
\includegraphics[width=0.66\columnwidth]{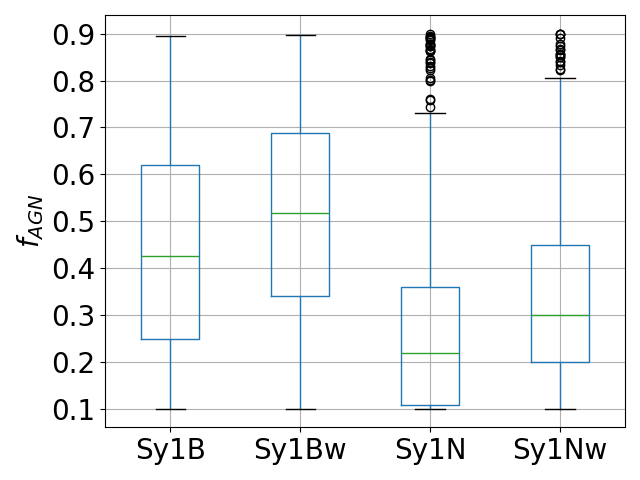}
\includegraphics[width=0.66\columnwidth]{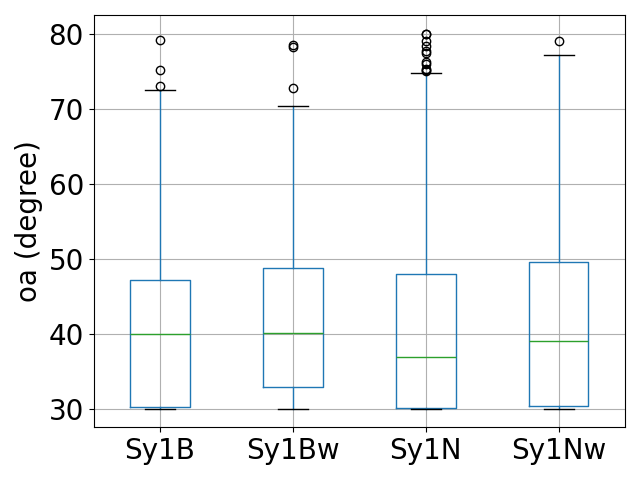}
\includegraphics[width=0.66\columnwidth]{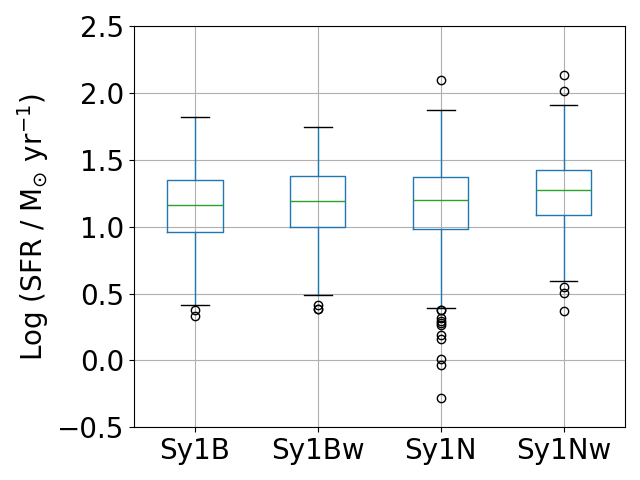}
\caption{Box-whisker plots for $f_{\textit{AGN}}$, \textit{oa}, \textit{SFR} for the Sy1 in the SW2M sample.}
\label{fig:SW2M_bwp}
\end{figure*}

The box-whisker plots for $f_{AGN}$, \textit{oa} and SFR are presented in Figure~\ref{fig:SW2M_bwp}, the medians are reported in Table~\ref{tab:SFR-fagn-i-oa_SW2M} and the results for the Dunn's post-test can be found in Table~\ref{tab:dunn-SFR-fagn-i-oa_SW2M}. For $f_{AGN}$ the box-whisker plots are similar to those obtained for the sample SW. More specifically, they show the same difference for $f_{AGN}$ to be higher in Sy1B than in Sy1N and to be higher in Sy1 with wind than in their counterparts without wind. All these differences are confirmed at the highest level of significance in Table~\ref{tab:dunn-SFR-fagn-i-oa_SW2M}, and by the median values of $f_{AGN}$ in Table~\ref{tab:SFR-fagn-i-oa_SW2M}, which are almost the same than in the SW sample. This implies that the addition of three data from 2MASS in NIR had no effect on the fits of this parameter. These differences in $f_{AGN}$ characterize the different spectral subgroups.

Like in the sample SW, the box-whisker plots for \textit{oa} in Figure~\ref{fig:SW2M_bwp} show the same trends: $oa$ is lower in Sy1N than in Sy1B, and higher in Sy1 with wind than without wind. However, in Table~\ref{tab:dunn-SFR-fagn-i-oa_SW2M}, the differences are less pronounced in the SW2M than in the SW sample. This also appears in the median values reported in Table~\ref{tab:SFR-fagn-i-oa_SW2M}. 

Comparing Figure~\ref{fig:SW2M_bwp} with Figure~\ref{fig:Sy1FR_BxPl} we do not distinguish much difference and comparing Table~\ref{tab:SFR-fagn-i-oa} with Table~\ref{tab:SFR-fagn-i-oa_SW2M} the medians for SFR in the Sy1B and Sy1Nw are exactly the same in the two samples, although lower for the Sy1Bw and Sy1N in the SW2M sample. In Table~\ref{tab:dunn-SFR-fagn-i-oa_SW2M}, the difference between Sy1N and Sy1Nw increases, while for any other pairs the results are the same as before. 
\begin{table}
\centering
\caption{Median values from the generic SED for the Sy1 in the SW2M sample.}
\label{tab:SFR-fagn-i-oa_SW2M}
\begin{tabular}{|c|c|c|c|c|}
\hline
Subgroup & $\log(\rm{SFR})$ & $f_{\rm AGN}$ & \textit{i}  &\textit{oa}   \\ 
&(M$_\odot$ yr$^{-1}$)&&(degree)&(degree)\\
\hline
Sy1B & 1.17 & 0.43 & 9.63 & 39.94 \\ 
Sy1Bw & 1.19 & 0.52 & 9.79  & 40.11 \\ 
 Sy1N & 1.19 & 0.22 & 9.53 & 36.81 \\ 
Sy1Nw & 1.27 & 0.30 & 9.74 & 39.07 \\ \hline
\end{tabular}               
\end{table}
\begin{table}
\centering
\begin{threeparttable}
\caption{Summaries of Dunn’s Multiple Comparisons Test for the Sy1 subgroups in the SW2M sample. Levels of significance as explained in Table 2.}
\label{tab:dunn-SFR-fagn-i-oa_SW2M}
\begin{tabular}{|c|c|c|c|c|}
\hline
Pairs & SFR & $f_\textit{AGN}$ & \textit{i}  &\textit{oa}   \\ 
\hline
Sy1B-Sy1Bw  & ns & **** & ns & * \\
Sy1B-Sy1N   & ns & **** & ns  & * \\ 
Sy1B-Sy1Nw  & **** & **** & ns & ns \\ 
Sy1Bw-Sy1N  & ns & **** & * & ** \\ 
Sy1Bw-Sy1Nw &** &****& ns & ns \\ 
Sy1N-Sy1Nw  & **** &**** & ns & * \\ 
\hline
\end{tabular}
\begin{tablenotes}
\item Notes: Levels of significance as explained in Table 2.
\end{tablenotes}
\end{threeparttable}
\end{table}
\begin{table}
\centering
\begin{tabular}{|c|c|c|c|c|}
\hline
\textbf{Subgroup} & \textbf{Parameter} & \textbf{Corr. coef.} & \textbf{Slope} & \textbf{Intercept} \\
\hline
\multirow{3}{*}{Sy1B} & $f_{AGN}$ & 0.93 &0.85 & 0.09 \\
                      & $oa$      & 0.89 & 0.88 & 4.37 \\
                      & $\log(\rm SFR)$ & 0.87 & 0.83 & 0.15 \\
\hline
\multirow{3}{*}{Sy1Bw} & $f_{AGN}$ & 0.92 & 0.83 & 0.11 \\
                       & $oa$      & 0.91 & 0.87 & 4.48 \\
                       & $\log(\rm SFR)$ & 0.83 & 0.74 & 0.26 \\
\hline
\multirow{3}{*}{Sy1N}  & $f_{AGN}$ & 0.90 & 0.88 & 0.08 \\
                       & $oa$      & 0.87 & 0.83 & 6.79 \\
                       & $\log(\rm SFR)$ & 0.89 & 0.86 & 0.08 \\
\hline
\multirow{3}{*}{Sy1Nw} & $f_{AGN}$ & 0.92 & 0.87 & 0.08 \\
                       & $oa$      & 0.89 & 0.85 & 5.81 \\
                       & $\log(\rm SFR)$ & 0.86 & 0.80 & 0.19 \\
\hline
\end{tabular}
\caption{Linear regressions for each spectral subgroup and parameter.}
\label{LR_SW-SW2M}
\end{table}

Although one might be tempted to attribute the cause of the lower significance of the statistical tests to possible physical differences in the SEDs, due to an addition of data in NIR, a comparison of the data in the two samples in Figure~\ref{fig:SW2MvsSW} of the Appendix rather suggests it is caused by a change in the dispersions of the data due to the reduction of the number of data close to $30$\% in the SW2M sample (see Table~\ref{tab:Sy1}). This can be better appreciated for $f_{AGN}$ in Figure~\ref{fig:SW2MvsSW}, where one can clearly distinguish an accumulation of points at low values in the Sy1N and Sy1Nw compared to Sy1B and Sy1Bw. This accumulation of points is what the statistical tests interpret as a significant difference. What is meaningful is not the individual values but their distributions which are characterized by their ranks. 

As judged by the coefficients of correlation reported in Table~\ref{LR_SW-SW2M} for the linear regressions comparing the data in SW with SW2M in Figure~\ref{fig:SW2MvsSW}, one can see that the variances (a divergence of results in the SEDs) is not the cause of the lowering of significance levels of the statistical tests in the SW2M sample. Except for the Sy1Bw, the linear fits are quite similar in all the spectral subgroups, with high correlation coefficients and only slightly flatter slopes. These slopes are consistent with the general trend noted before for the values in SW2M to be slightly higher than in SW. These higher values are consistent with an observational bias affecting 2MASS data.

In the SW sample the SFR was found to decrease along the sequence Sy1Nw$\rightarrow$Sy1N$\rightarrow$Sy1Bw$\rightarrow$Sy1B. In the SW2M the medians in SFR being the same as before in the Sy1B and Sy1Nw, but decreasing in the Sy1N and Sy1Bw, the sequence in SFR would have changed to Sy1Nw$\rightarrow$Sy1N/Sy1Bw$\rightarrow$Sy1,due only to a change in dispersion of SFRs in the intermediate Sy1N and Sy1Bw. This suggests that the trend for Sy1N to have higher SFR than Sy1B is valid in both samples. Moreover, considering that the difference in SFR is lower between the Sy1Nw and Sy1Bw than between the Sy1Nw and the two other Sy1 without wind, then we could also conclude that the trend for Sy1 with wind to have higher SFR than those without wind, as confirmed in the SW sample, is still observed in the SW2M sample. 

In general, therefore, we conclude that adding the three fluxes of 2MASS in X-CIGALE did not affect the results obtained using only the WISE fluxes, and that the main reason why the differences in $oa$ and SFR are less statistically significant in the SW2M sample is solely due to the reduction by almost 30\% of the data in the SW2M sample, which makes the detection of accumulations at some characteristics values consistent with differences in parameters between the spectral subgroups less obvious. 

We therefore accept as robust the results for X-CIGALE analysis that suggests 1- $i$ is small, consistent with UPAGN, 2- $f_{AGN}$ is a characteristic of the SED that explains in part the spectral subgroups, 3- the covering factor, $oa$, is larger in Sy1 with wind than in Sy1 without wind, and possibly also larger in Sy1B than in Sy1N, and 4- SFR is starburst like, and possibly higher in Sy1N/Sy1Nw than in Sy1B/Sy1Bw.

\section{Discussion} 
\label{discussion}

The generic SED consistent with UPAGN implies that all the Sy1 in our sample have a small LOS angle, $i$, close to $10^\circ$, which means that we have a direct view of the broad line region (BLR) and accretion disk (although this disk is not resolved). However, at the same time, this result eliminates the intermediate angle hypothesis: the appearance of narrow lines in the spectra of the Sy1N is not due to an intermediate LOS angle relative to the torus. To explain the spectral subgroups we see now only two alternatives, which are that either Sy1B and Sy1N have different tori or their BLRs are different.

The next best hypothesis consistent with UPAGN is a difference in the level of ``clumpiness'' of the torus \citep{2002Nenkova,2007Tristram,2011RamosAlmeida, 2015RamosAlmeida,2016Mateos,2017Bisogni,2017Audibert}. More specifically, a torus punctured by a higher number of holes allows more light from the AGN to reach the disk, exciting the gas in the NLR. In the model of torus we used \citep{2012Stalevski}, the half-opening angle, $oa$, is equal to the covering factor in the optical, $f_{cov}$. Consequently, lower $oa$ means smaller $f_{cov}$ consistent with more clumpy torus. However, this explanation would also need to point to a critical level of clumpiness above which no light can pass through the torus in the Sy1B, and, hopefully also, reveals evidence why the tori are different. 

To explore this question using X-CIGALE, we added four more free parameters in the model of the torus that might affect it. These are, $t$, the edge-on optical depth at 9.7 $\mu$m, $pl$, a power law for the radial dust density, $q$, a density gradient varying with the polar angle, and $R$, the ratio of outer to inner radius of the torus. To distinguish which of the SED models is better, we used the BIC.  All the modified SED models produce significantly higher BIC than the generic model, clearly favouring the latter. Moreover, all the modified SED models predict differences in SFR, $f_{AGN}$ and $i$, exactly as in the generic model, except for $oa$, where no differences is detected. However, none of the new torus parameters in these models seem to yield any clue about why this could be happening. Consequently, we feel safe in rejecting the hypothesis that a difference in tori explains the spectral subgroups.   

The second alternative, the BLRs in the spectra subgroups are different, looks more probable. More specifically, a richer (more extended or massive) BLR in  Sy1B than Sy1N produces more intense broad emission lines, hiding any narrow lines in their spectra. Assuming matter in the BLR feeds the accretion disk \citep{2009Gaskell}, this hypothesis would agree with all the other differences observed between the subgroups \citep[see Table~4 in][]{2020Torres-Papaqui}: Sy1B have higher bolometric luminosities, steeper power laws in optical, and despite having more massive BHs, higher Eddington ratios (${\rm L}_{bol}$⁄${\rm M}_{BH}$) than Sy1N. This may also explain the difference in $oa$: according to the model of \citet{2009Gaskell}, based on reverberation analyses, the BLR has the same $f_{cov}$ as the torus (see Figure~\ref{fig:model-oa}) and thus Sy1N having less massive or extended BLR than Sy1B must also have lower $f_{cov}$, i.g., lower $oa$.

Adopting the model of \citet{2009Gaskell}, another observation made by \citet{2020Torres-Papaqui} can be explained, which is the fact that  H$\beta$ FWHM in Sy1 with wind is systematically smaller than in Sy1 without wind. If the gas from the BLR replenishes the accretion disk, a sudden increase in accretion in Sy1 with wind--in fact, triggering the wind--depletes the BLR in gas, explaining the smaller FWHMs. This explanation also agrees with the observational evidence of higher accretion rates \citep{2020Torres-Papaqui}: Sy1 with wind have higher bolometric luminosities and Eddington ratios than their counterparts without wind.  

According to the generic SED model, therefore, a difference in BLR seems to be the cause of the spectral subgroups. But the generic model also suggests $oa$ is systematically larger in Sy1 with wind than in their counterparts without wind. When we started our analysis, we thought this could be an artifact of X-CIGALE, the program confounding the wind with the torus. However, another model based on hydrodynamic simulations was suggested to us by the reviewer that can better explain this result. This is the fountain model as proposed by \citet{2012Wada,2015Wada}. According to this model, matter from the wind replenishes the torus, explaining the larger $oa$. The fountain model also predicts the wind phenomenon is recurrent \citep{2014Schartmann}, matter from the torus eventually falling back to the accretion disk, triggering new wind events. This model also suggests something even more important, which is that AGN winds might have built the torus \citep[for an alternative see][and references therein]{2012Hopkins}. 

However, we see two possible difficulties with the fountain hypothesis as applied to the Sy1. The first is that the winds detected by \citet{2020Torres-Papaqui} are structures located in the NLR. They are outflow (OF) components in the [OIII] line at 5007 \AA, at kpc from the central engine, while the fountain model covers only a few 10s of pc, which is the scale of the torus. Consequently, although it is easy to imagine matter ejected by AGN wind from the torus to fall back on the BLR (a sub-pc structure), it is much more difficult to imagine matter from the NLR to first fall back on the torus, then from the torus to the BLR and finally on the accretion disk. However, before reaching the NLR, the wind must definitely pass by the region of the torus and thus it is highly conceivable to assume that part of the matter from the wind replenishes the torus \citep{2020Williamson,2016Wada} making the process recurrent \citep[for evidence of recurrence in AGN see][and reference therein]{2025Lyu}.

The second difficulty is that assuming recurrence one could have expected to find evidence of previous wind events in the NLR of AGN without wind, which is not the case of Sy1 where no OF is detected (Sy1B and Sy1N). This might be because evidence of previous wind event in the disk of AGN hosts might not live long, especially if the wind has a positive feedback effect, like triggering star formation. However, in our study, enhanced star formation in Sy1 with wind only appears as weak trends in Table~\ref{tab:dunn-SFR-fagn-i-oa}, the  significant difference being higher SFR in Sy1N than in Sy1B. Moreover, in \citet{2020Torres-Papaqui} no correlation was found between the parameters related to the SMBH or the wind with the SFR. On the other hand, if recurrent AGN winds are responsible in building the torus, this structure itself could be taken as evidence of past wind events in Sy1 now without wind. 

\subsection{Connecting UPAGN to the formation of galaxies}

In Sy1 \citep{2020Torres-Papaqui} and in other AGN types \citep{JP2024} no evidence was found of AGN winds quenching star formation. In fact, these two analyses revealed nothing about the possible effect AGN winds have or have had on their host galaxies. Thanks to X-CIGALE, we now have additional information consistent with UPAGN about the putative torus in Sy1 which could shed new light on this crucial question: what role does AGN winds play in the formation of galaxies? 

Two "new" aspects of the accretion process in Sy1 seems to appear in our present study: 1- AGN wind is triggered by a sudden accretion of matter from the BLR, and 2- AGN wind has a direct impact on the torus, replenishing it in matter; then matter from the torus can fall back on the BLRG and eventually on the accretion disk, making the phenomenon recurrent. However, so far, nothing in our analysis tells us why only a fraction of the Sy1B and Sy1N are seen to develop a wind in the first place.

\begin{figure*}
\includegraphics[width=0.9\columnwidth]{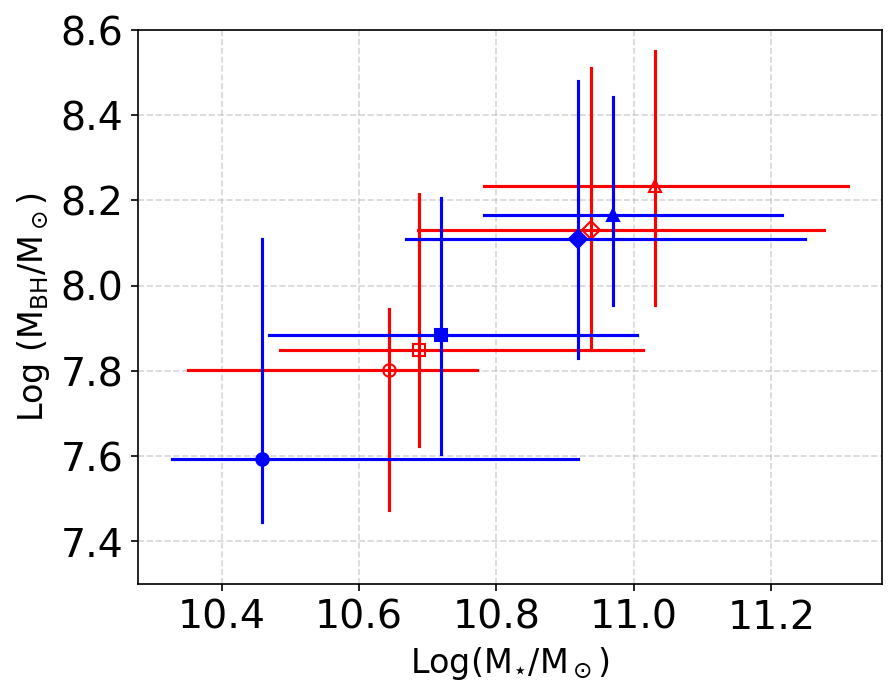}
\includegraphics[width=0.9\columnwidth]{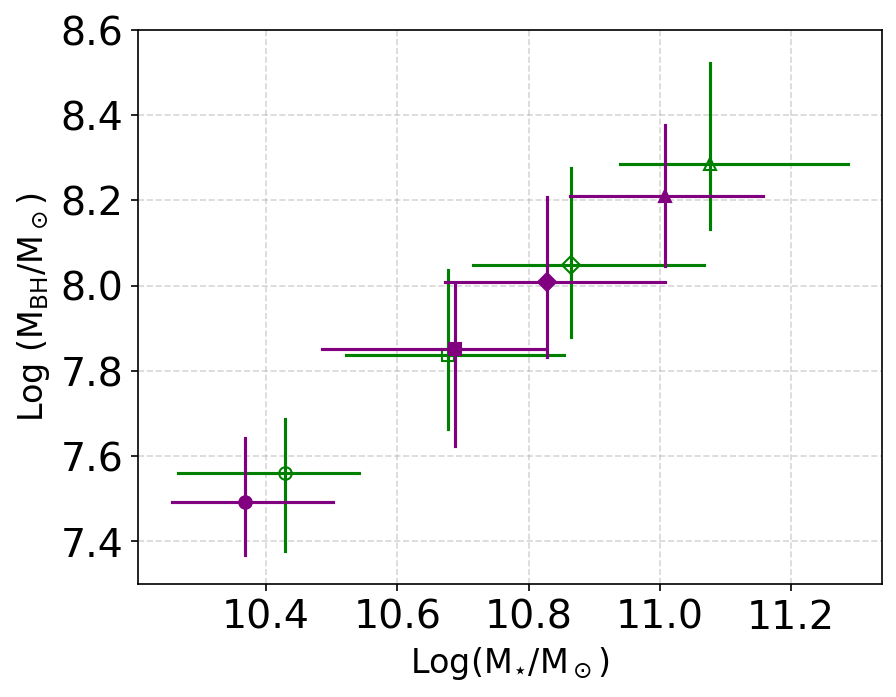}
\caption{Comparing the mass of Sy1 galaxy hosts with and without wind in four intervals of redshift bins: circle, $z$ from (0, 0.1], square, (0.1, 0.2], lozenge, (0.2, 0.3] and triangle, (0.3, 0.4]. In each figure the points are the medians and the bars are the quartiles Q1 and Q3. On the left panel, Sy1B (red open symbol) and Sy1Bw (blue filled symbol), on the right, Sy1N (green open symbol) and Sy1Nw (magenta filled symbol).}
\label{fig:Mbh_vs_Mestellar}
\end{figure*}

\begin{figure*}
\includegraphics[width=0.9\columnwidth]{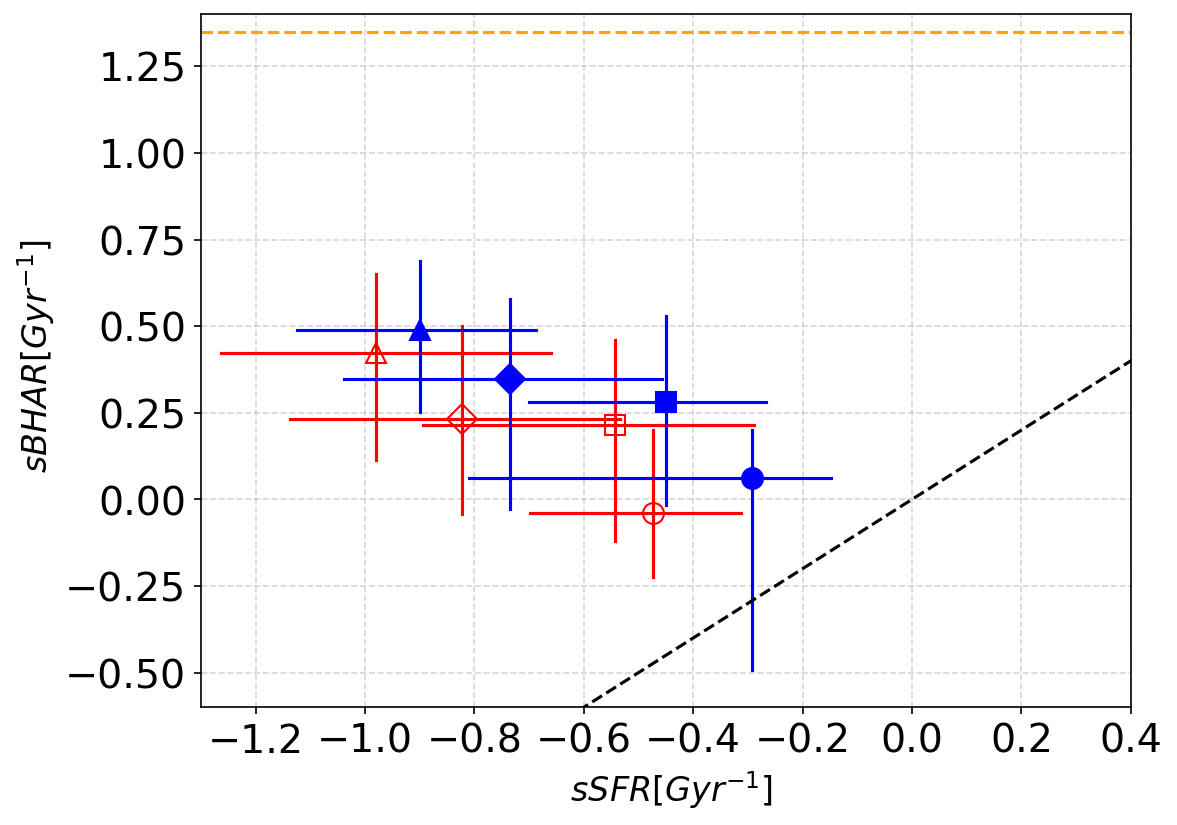}
\includegraphics[width=0.9\columnwidth]{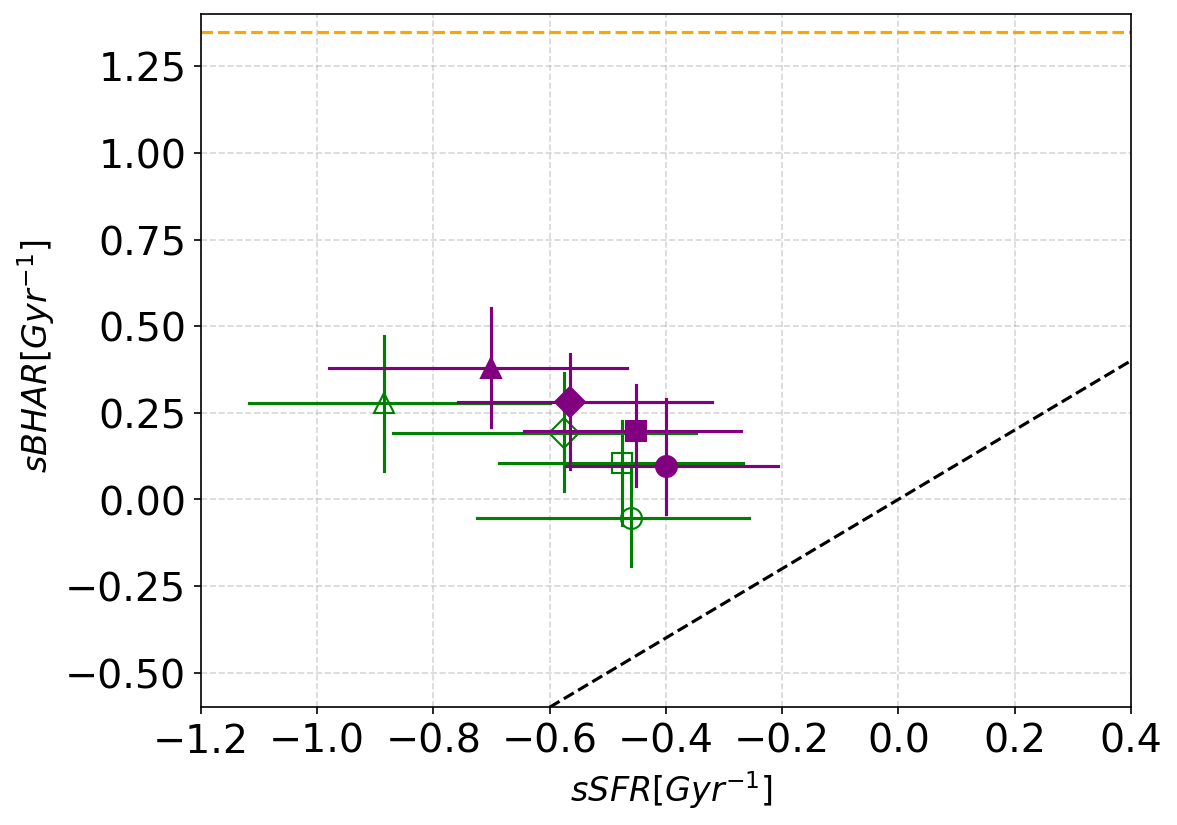}
\caption{Specific diagnostic diagrams for the Sy1B on the left and Sy1N on the right. The points, bars and colors are as in Figure~\ref{fig:Mbh_vs_Mestellar}. The black dashed diagonal line is the one-to-one relation, where a galaxy and its SMBH grow in mass at the same rate; the farther to the left of the diagonal the faster the BH grow compared to the galaxy. The yellow horizontal dashed line is the Eddington limit in specific accretion onto the BH.}
\label{fig:sBHAR_vs_sSFR}
\end{figure*}

One possible answer is evolution: AGN with wind are at an earlier phase of formation than their counterparts without wind. In \citet{JP2024}, a scenario was proposed based on the specific diagnostic diagram, comparing the specific accretion rate (sBHAR; the accretion rate divided by the BH mass) with the specific star formation rate (sSFR; the star formation rated divided by the galaxy host mass), where the evolution is the same in all AGN, independently of their activity types (RG, LINER, Sy2 or Sy1), and where AGN with wind only evolve into AGN of the same type without wind, because the AGN types correspond to different formation processes of the host galaxies \citep[see explanations in][]{JP2024}. Assuming AGN winds are ubiquitous \citep{2020Torres-Papaqui} and are responsible in forming the tori \citep{2012Wada}, this evolution would right away agrees with UPAGN, since within this paradigm all AGN must have such a structure. This also straightforwardly solves the problem of finding evidence for the effect of AGN wind \citep[][]{2020Torres-Papaqui,JP2024}, since the mere presence of a torus is such evidence. 

Therefore, the next natural step in our analysis consists in comparing the locus of the Sy1 subgroups in the specific diagnostic diagram to verify  how well they could fit the evolution hypothesis. Because X-CIGALE produces higher SFRs than previously estimated in \citet{2020Torres-Papaqui,JP2024} we need first to recalculate sSFR. What is missing to do this is the stellar masses, $M^*$, of the host galaxies. We did not use the masses produced by X-CIGALE because although the range in $M^*$ are comparable to what was estimated in the previous two studies, the values do not yield a relation with the BH mass, M$_{BH}$--M$^*$, and all the studies in the literature predicts such a relation \citep[ex.][]{2004Haring,2015Reines,2022Ramsden}. After comparing these different relations we chose the one determined by \citet{2004Haring}, which seems to yield intermediate values compared to \citet{2015Reines} and \citet{JP2024}.

In Figure~\ref{fig:Mbh_vs_Mestellar}, M$^*$ is compared with M$_{BH}$, separating the sample in four redshift bins. In general, Sy1 with wind seems to have lower masses than Sy1 without wind. This is a trend observed in all AGN, independently of their classification in activity types \citep{JP2024}. However, the trend in Sy1 is particularly weak (weaker than in any other AGN type), and distinguishing between Sy1N and Sy1B we see the contrary in the second bins. This is possibly due to downsizing, because the masses of the SMBH (and host galaxy) increases with the redshift and Sy1 with wind are usually at slightly higher redshifts than their counterparts without wind (cf. Figure~\ref{fig:z-Sy1}). 


The most relevant differences between AGN with and without wind really appear in the specific diagram, where sSFR and sBHAR are significantly higher in the former than in the latter; this is a common trait of all AGN with wind \citep{JP2024}. This particularity of Sy1 with wind is easily distinguishable in Figure~\ref{fig:sBHAR_vs_sSFR} comparing the different spectral subgroups in different redshift bins. Because SFRs estimated using X-CIGALE are higher than in the previous studies, Sy1 are now closer to the one-to-one relation, which is the locus where the host galaxies are growing in mass at almost the same rate as their SMBHs. This implies that we observe all these galaxies during an important phase of their formation, where, because they are more active, Sy1 with wind are less evolved than their counterparts without wind. 

Within the evolutionary scenario for the Sy1 with wind, we would naturally expect to see sSFR and sBHAR decreasing with time, as more gas from the disk disappear into stars and less gas falls from the torus to the BLR, and from the BLR to the accretion disk, the conditions for the formation of wind becoming less favourable. With time, therefore, Sy1 with winds would become similar to Sy1 without wind, leaving behind massive tori as evidence of the higher AGN activity in their past.   

In Figure~\ref{fig:sBHAR_vs_sSFR}, we also distinguish a decrease of sSFR as sBHAR increases at higher redshifts. This could be another evidence of downsizing: more massive galaxies at high redshifts accreting more matter onto their SMBHs (higher sBHAR) and forming more stars rapidly (having lower sSFR), losing their gas more rapidly, tend to be farther from the one-to-one relation at higher redshifts. Because the same behaviour is observed in all AGN at low redshifts independently of their activity type \citep{JP2024}, this suggests that, except for downsizing effect, the formation processes and evolution of all these galaxies were fundamentally the same.  
  
Because SFRs in Sy1 are starburst-like, as in QSOs \citep{2023Cutiva-Alvarez}, and because Sy1 galaxies are frequently considered to be the low luminosity equivalent of QSOs at low redshifts \citep{2006Osterbrock}, one might  have expected to observe the same pattern of evolution for both objects in the specific diagnostic diagram. However, this is not the case, since the specific diagnostic diagram of QSOs \citep[Figure~14 in][]{2023Cutiva-Alvarez} shows that over a large range in $z$, from $z\sim 0.25$ to $z\sim 4$,  both sSFR and sBHAR decrease with the redshift, tracing a sequence parallel to the one-to-one relation, while in Sy1 the sSFR increases as sBHAR decreases with the redshift (the sequence being almost perpendicular to the one-to-one relation). 

This difference can be explained assuming galaxies at high $z$ (QSOs) form in denser environments than galaxies at low $z$ (Sy1) in lower density environments. More specifically, this has to do with the rapidity with which the SMBH form at high redshifts: at high $z$ more gas is falling on the SMBH of QSOs than at low $z$ in the Sy1, and the more massive the SMBH, the faster the formation of the galaxy hosts (forming massive bulges), and thus the faster the depletion of gas, moving the sSFR in the specific diagnostic diagram farther from the one-to-one relation. The parallel sequence of QSOs, therefore, is solely due to downsizing.  

The differences in locus of the QSOs and Sy1 in the specific diagram would thus be due to different time-scales in the processes of galaxy host formations in different environments, the formation of SMBH and their host galaxies being much less rapid in the latter (spiral galaxies in the field) than in the former (galaxies in rich groups or clusters). 

A rapid depletion of gas due to a rapid formation of the SMBHs and galaxy hosts at high redshifts also explains why in AGN star formation seemed to be quenched. In all our studies of AGN with OF, we saw no evidence of quenching by the wind, which unique role, at least in Sy1 (and possibly Sy2), is to gradually build the torus, allowing mass to eventually falls back on the BLR, prolonging the AGN activity lifetime. At very high redshifts, the rapid formation of SMBH pushed to the extreme might explain why we see very massive SMBHs in apparently unusually lightweight host galaxies, the formation of the bulge being extremely fast and very early on most of the mass ends into the SMBH \citep{2023Boylan-Kolchin,2023Labbe,2024AJChworowsky,2024Tripodi}.

\section{Conclusions}

Our analysis with X-CIGALE shows that it is possible to construct a generic SED for Sy1 galaxies that is fully consistent with the unification paradigm of AGN (UPAGN): 90\% of the Sy1 galaxies in our sample have a small LOS angle, $i \sim 10^\circ$, consistent with a direct and open view of the central engine. 

This result has two immediate consequences: 1- the Sy1N spectral subgroups cannot be explained by the torus being seen at intermediate LOS angles, and 2) the detection by X-CIGALE of a torus in the SED of Sy1, where no evidence of obscuration exists, is another evidence in favour of UPAGN, according to which all AGN must form such a structure. 

Our analysis with X-CIGALE also suggests two important new facts about the relation of the torus with the activity of the SMBH: 1- following \cite{2009Gaskell}, the mass of gas in the torus falls back on the BLR, and from there to the accretion disk, producing a sudden increase in accretion responsible in triggering the wind, 2) following \citet{2012Wada}, matter ejected by the wind replenishes the torus, making the production of AGN wind a recurrent event.

Combining the two models we arrived to the conclusion that Sy1 with wind are at an earlier phase of evolution, eventually transforming into Sy1 without wind within the same spectral subgroups, since the origin of these subgroups is related to different formation processes of their host galaxies \citep{JP2024}. During this phase, AGN winds build the torus, this structure remaining as evidence of previous higher level of activity in Sy1 without wind. 

According to our study, what we observe in Sy1 is common to all AGN galaxies, irrespective of their activity types \citep{JP2024}. This is another argument in favour of UPAGN. As a final test, therefore, we will do a similar analysis in Part~II by applying the generic model to a large sample of Sy2 (18,585 galaxies).

\section*{Acknowledgements}
\addcontentsline{toc}{section}{Acknowledgements}

R-A.P.A., C.R. and T-P.J.P. would like to thanks the reviewer, Dr. Keiichi Wada for suggesting that our results could be consistent with the fountain model and the formation of the torus by AGN winds, and for the multiple comments that allowed us to make our study clearer and to the point. R-A.P.A. acknowledges SECIHTI for its support through grant CVU  1245642. For his part, T-P.J.P. and C.R. acknowledge DAIP-UGTO (Mexico) for grant support 0077/2021.

This research has used the VizieR catalogue access tool, CDS, Strasbourg, France (DOI : 10.26093/cds/vizier). The original description of the VizieR service was published in 2000, A\&AS 143, 23. Funding for SDSS-III has been provided by the Alfred P. Sloan Foundation, the Participating Institutions, the National Science Foundation, and the U.S. Department of Energy Office of Science. The SDSS-III web site is http://www.sdss3.org/.

\section*{Data Availability}
The data underlying this article will be shared on reasonable request to the corresponding author.

\bibliographystyle{mnras}
\bibliography{bibliografia}


\begin{appendices}
\section{Comparison of the results for the sample SW with those for the sample SW2M}

\begin{figure*}
\includegraphics[width=0.66\columnwidth]{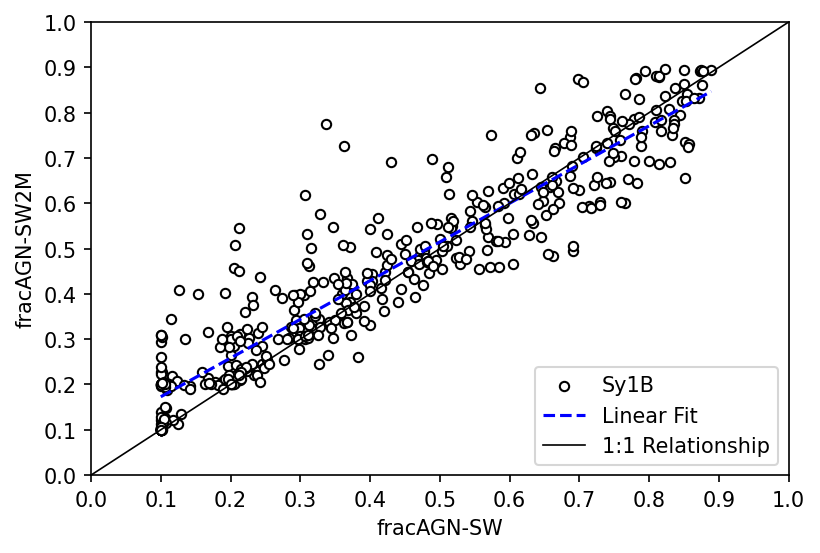}
\includegraphics[width=0.66\columnwidth]{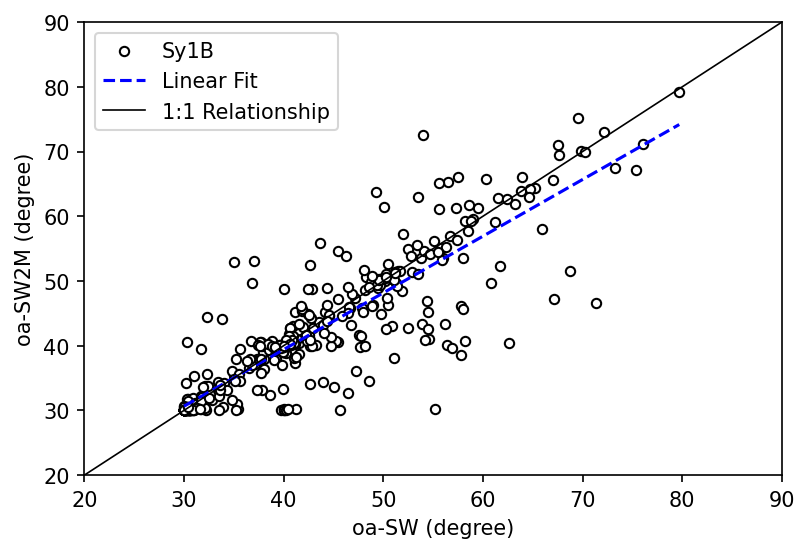}
\includegraphics[width=0.66\columnwidth]{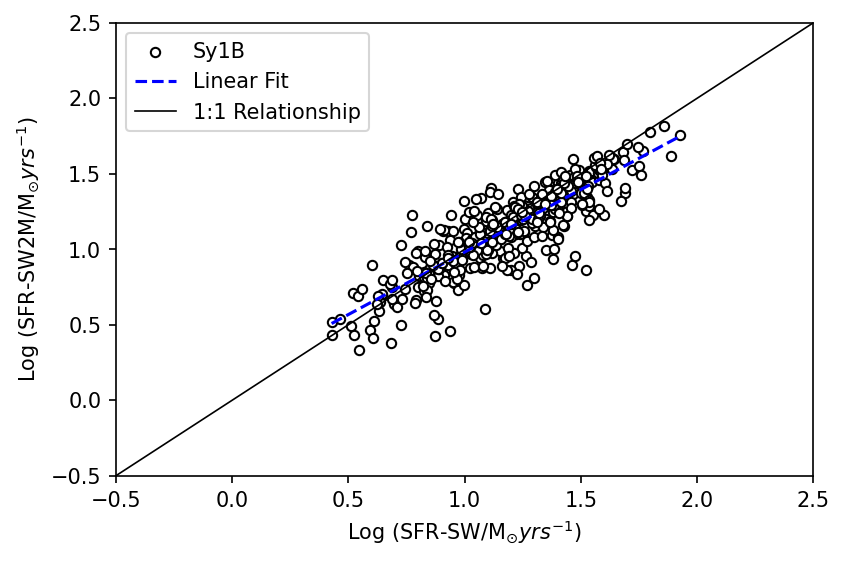}
\includegraphics[width=0.66\columnwidth]{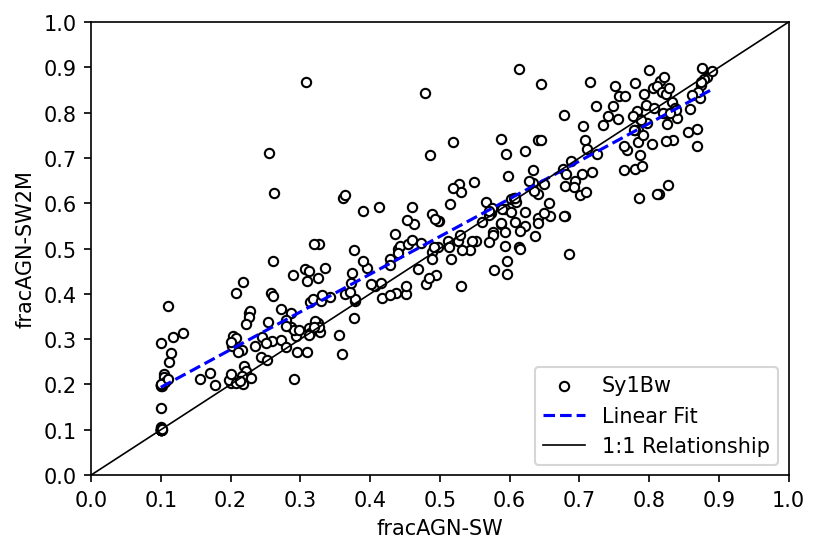}
\includegraphics[width=0.66\columnwidth]{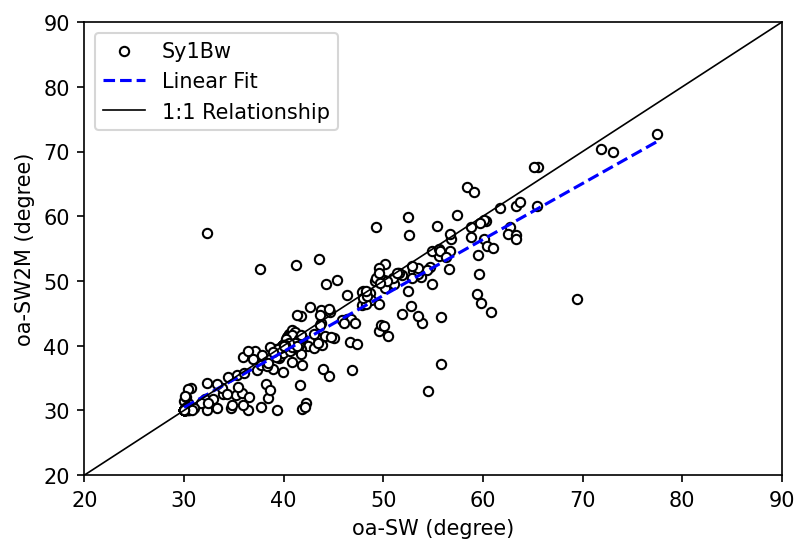}
\includegraphics[width=0.66\columnwidth]{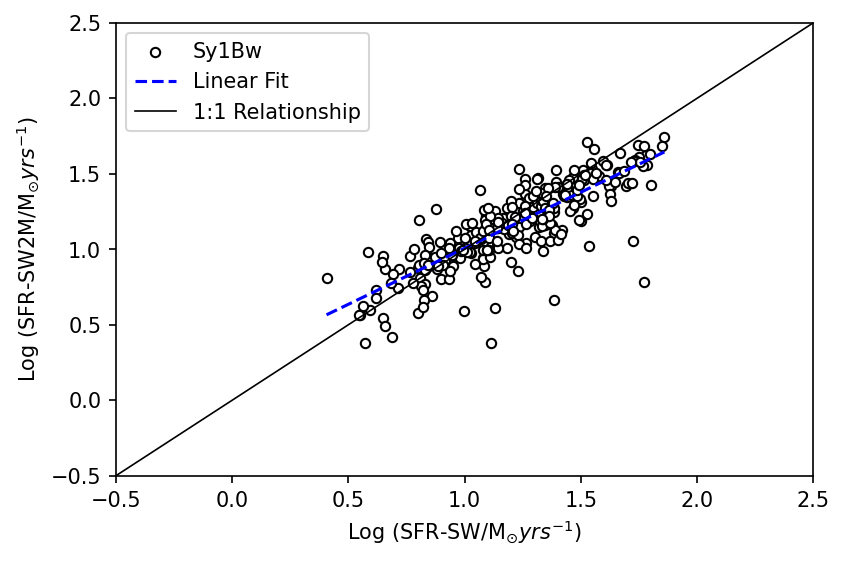}
\includegraphics[width=0.66\columnwidth]{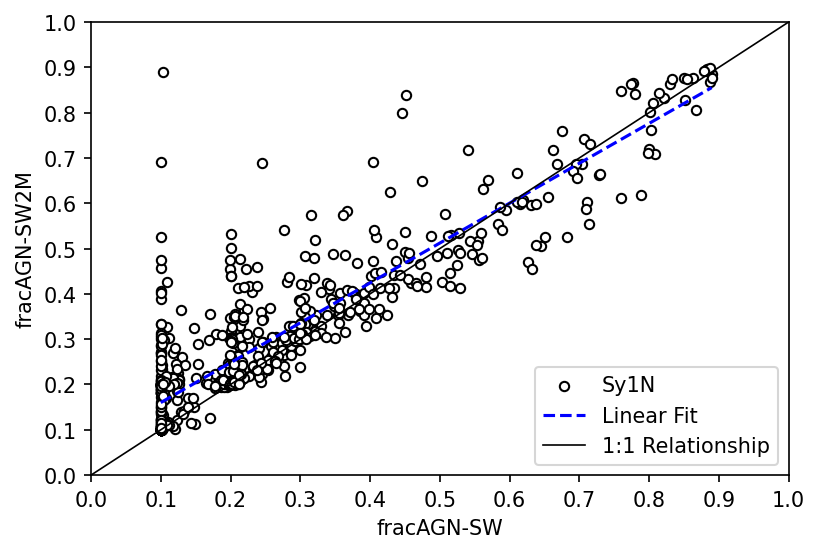}
\includegraphics[width=0.66\columnwidth]{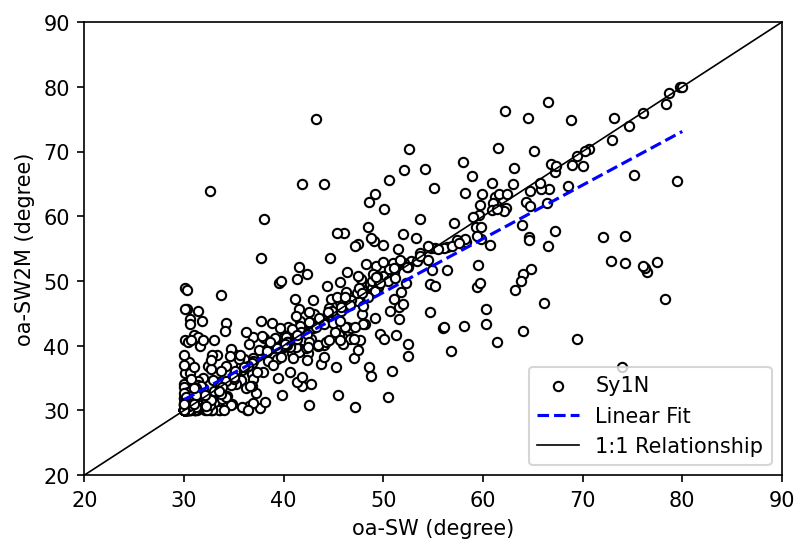}
\includegraphics[width=0.66\columnwidth]{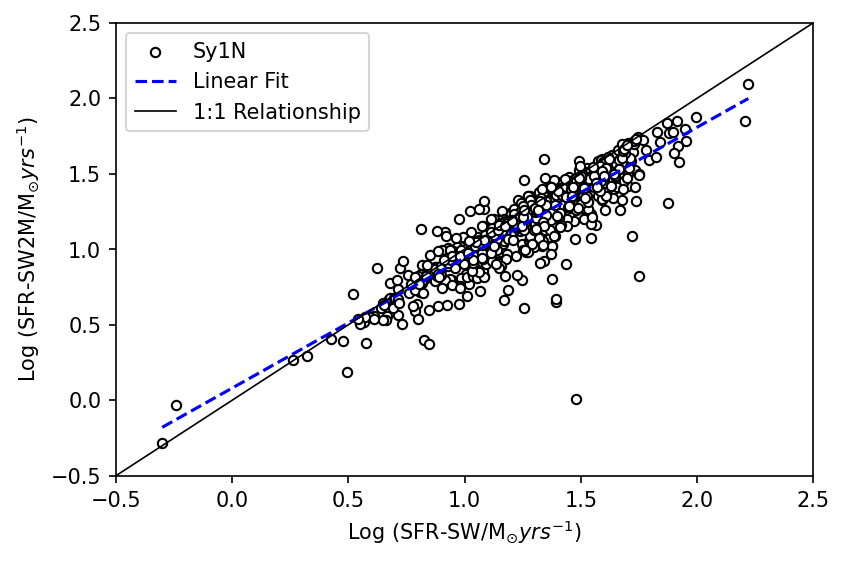}
\includegraphics[width=0.66\columnwidth]{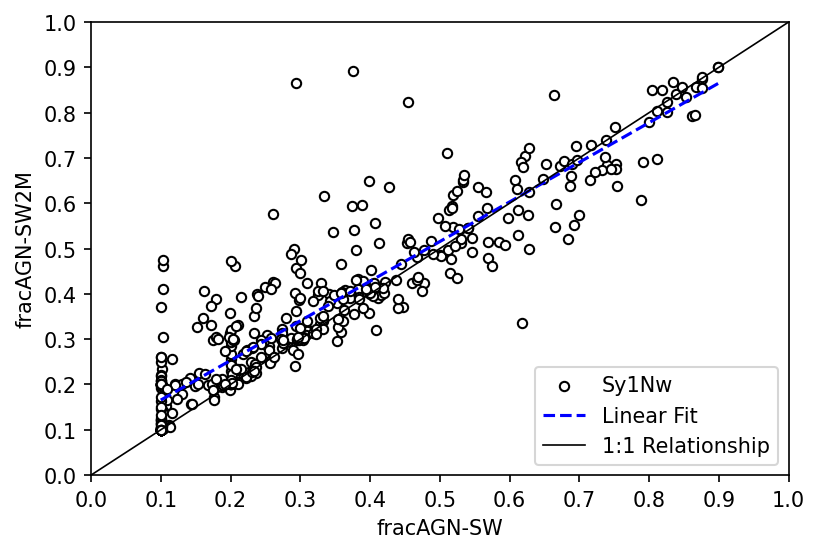}
\includegraphics[width=0.66\columnwidth]{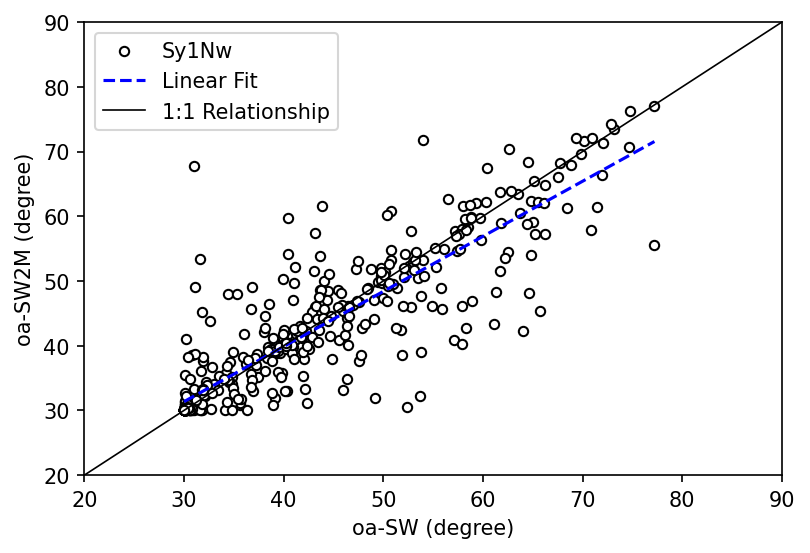}
\includegraphics[width=0.66\columnwidth]{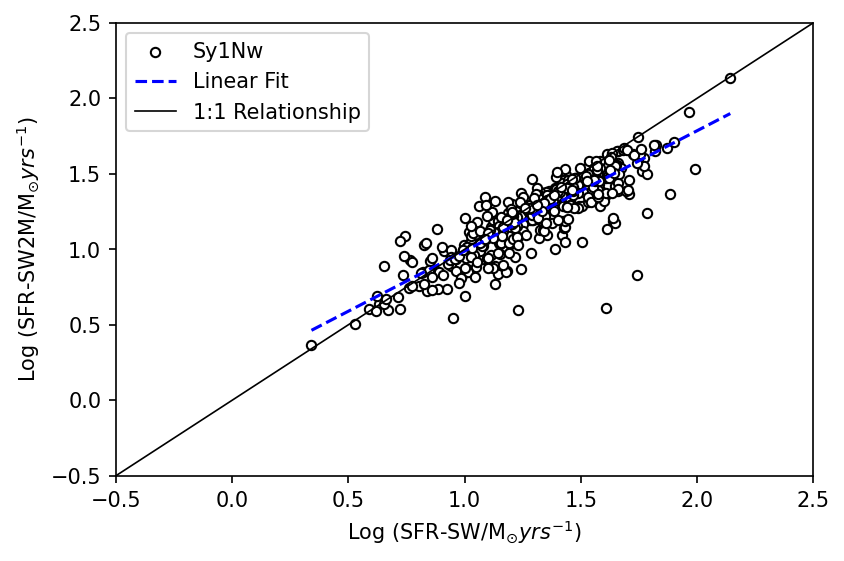}
\caption{Comparing the parameters $f_{AGN}$, $oa$ and SFR, in the SW2M and SW samples. A one-to-one relation is traced (continuous black line) while the dashed blue lines are linear regressions.}
\label{fig:SW2MvsSW}
\end{figure*}

In Figure~\ref{fig:SW2MvsSW} we compare the results for $f_{AGN}$, $oa$ and SFR as obtained using the Sy1 generic model on the data in the SW2M and SW samples. In each panel, a one-to-one relation is traced as a black continuous line, and a linear regression as a blue dashed line. The parameters of the regressions are reported in Table~\ref{LR_SW-SW2M} and discussed in the text in Section~\ref{Results}. 

\end{appendices}

\end{document}